%Paper: hep-ph/9503445
%From: Stephen Martin <spmartin@umich.edu>
%Date: Mon, 27 Mar 1995 13:14:47 -0500 (EST)
%Date (revised): Wed, 26 Apr 1995 12:26:57 -0400 (EDT)

%%%%%%%%%%%%%%%%%%%%%%%%%%%%%%%%%%%%%%%%%%%%%%%%%%%%%%%%%%%%%%%%%%%%%%%%%
%  Low-energy supersymmetry with D-term contributions to scalar masses  %
%                            by                                         %
%             Chris Kolda and Stephen P. Martin                         %
%                  University of Michigan                               %
%                28 pages, plain TeX, plus 12 figures                   %
% guide to revisions: paragraph surrounding equations (2.11) and (2.12) %
% and the following two paragraphs; paragraph following equation (3.2); %
% and figs 6 and 8 have all been modified slightly.                     %
%%%%%%%%%%%%%%%%%%%%%%%%%%%%%%%%%%%%%%%%%%%%%%%%%%%%%%%%%%%%%%%%%%%%%%%%%

%  Define pseudo-12pt fonts
\font\twelverm=cmr10 scaled 1200    \font\twelvei=cmmi10 scaled 1200
\font\twelvesy=cmsy10 scaled 1200   \font\twelveex=cmex10 scaled 1200
\font\twelvebf=cmbx10 scaled 1200   \font\twelvesl=cmsl10 scaled 1200
\font\twelvett=cmtt10 scaled 1200   \font\twelveit=cmti10 scaled 1200
\skewchar\twelvei='177   \skewchar\twelvesy='60
%  Define \...point macros to change fonts and spacings consistently
\def\twelvepoint{\normalbaselineskip=12.4pt
  \abovedisplayskip 12.4pt plus 3pt minus 9pt
  \belowdisplayskip 12.4pt plus 3pt minus 9pt
  \abovedisplayshortskip 0pt plus 3pt
  \belowdisplayshortskip 7.2pt plus 3pt minus 4pt
  \smallskipamount=3.6pt plus1.2pt minus1.2pt
  \medskipamount=7.2pt plus2.4pt minus2.4pt
  \bigskipamount=14.4pt plus4.8pt minus4.8pt
  \def\rm{\fam0\twelverm}          \def\it{\fam\itfam\twelveit}%
  \def\sl{\fam\slfam\twelvesl}     \def\bf{\fam\bffam\twelvebf}%
  \def\mit{\fam 1}                 \def\cal{\fam 2}%
  \def\tt{\twelvett}
  \textfont0=\twelverm   \scriptfont0=\tenrm   \scriptscriptfont0=\sevenrm
  \textfont1=\twelvei    \scriptfont1=\teni    \scriptscriptfont1=\seveni
  \textfont2=\twelvesy   \scriptfont2=\tensy   \scriptscriptfont2=\sevensy
  \textfont3=\twelveex   \scriptfont3=\twelveex  \scriptscriptfont3=\twelveex
  \textfont\itfam=\twelveit
  \textfont\slfam=\twelvesl
  \textfont\bffam=\twelvebf \scriptfont\bffam=\tenbf
  \scriptscriptfont\bffam=\sevenbf
  \normalbaselines\rm}
%       tenpoint

%      Various internal macros
\def\beginlinemode{\endmode
  \begingroup\parskip=0pt \obeylines\def\\{\par}\def\endmode{\par\endgroup}}
\def\beginparmode{\endmode
  \begingroup \def\endmode{\par\endgroup}}
\let\endmode=\par
{\obeylines\gdef\
{}}
\def\singlespace{\baselineskip=\normalbaselineskip}
\def\oneandthreefifthsspace{\baselineskip=\normalbaselineskip
  \multiply\baselineskip by 8 \divide\baselineskip by 5}

\def\oneandahalfspace{\baselineskip=\normalbaselineskip
  \multiply\baselineskip by 3 \divide\baselineskip by 2}
\def\doublespace{\baselineskip=\normalbaselineskip \multiply\baselineskip by 2}
\newcount\firstpageno
\firstpageno=2
\footline={
\ifnum\pageno<\firstpageno{\hfil}\else{\hfil\twelverm\folio\hfil}\fi}
\let\rawfootnote=\footnote              % We must set the footnote style
\def\footnote#1#2{{\rm\singlespace\parindent=0pt\rawfootnote{#1}{#2}}}
\def\raggedcenter{\leftskip=2em plus 12em \rightskip=\leftskip
  \parindent=0pt \parfillskip=0pt \spaceskip=.3333em \xspaceskip=.5em
  \pretolerance=9999 \tolerance=9999
  \hyphenpenalty=9999 \exhyphenpenalty=9999 }
\parskip=\medskipamount
\twelvepoint            % selects twelvepoint fonts (cf. \tenpoint)
\overfullrule=0pt       % delete the nasty little black boxes for overfull box
\def\preprintno#1{
 \rightline{\rm #1}}    % Preprint number at upper right of title page
\def\author                     %  Author(s) name(s)  on title page
  {\vskip 3pt plus 0.2fill \beginlinemode
   \singlespace \raggedcenter \twelvesc}
\def\affil                      % Affiliations (can intermix with \author)
  {\vskip 3pt plus 0.1fill \beginlinemode
   \oneandahalfspace \raggedcenter \sl}
\def\abstract                   % Begin abstract
  {\vskip 3pt plus 0.3fill \beginparmode
   \doublespace \narrower \noindent ABSTRACT: }
\def\endtitlepage               % End title page, begin body of paper
  {\endpage                     %       This subsumes \body
   \body}
\def\body                       % Begin text body;  can be used to end
  {\beginparmode}               % \title, \author, \affil, \abstract,
                                % \reference, or \figurecaption modes

\def\subhead#1{                 % Subhead;  NOTE enclose the text in {}
  \vskip 0.25truein             % e.g., \subhead{A. History of the Problem}
  {\raggedcenter #1 \par}
   \nobreak\vskip 0.1truein\nobreak}
\def\refto#1{$|{#1}$}           % For references in text as superscript
\def\references                 % Begin references -- basic format is Phys Rev
  {\subhead{References}         % I.e., volume, page, year
   \beginparmode
   \frenchspacing \parindent=0pt \leftskip=1truecm
   \parskip=8pt plus 3pt \everypar{\hangindent=\parindent}}
\gdef\refis#1{\indent\hbox to 0pt{\hss#1.~}}    % Ref list numbers.
\gdef\journal#1, #2, #3, 1#4#5#6{               % Journal reference.  Comma
    {\sl #1~}{\bf #2}, #3, (1#4#5#6)}           % off: name, vol, page, year
\def\refstylenp{                % Nucl Phys(or Phys Lett) ref style: V, Y, P
  \gdef\refto##1{ [##1]}                                % Reference in text []
  \gdef\refis##1{\indent\hbox to 0pt{\hss##1)~}}        % Ref list numbers)
  \gdef\journal##1, ##2, ##3, ##4 {                     % Journal reference
     {\sl ##1~}{\bf ##2~}(##3) ##4 }}
\def\refstyleprnp{              % Input like pr, output like np!!
  \gdef\refto##1{ [##1]}                                % Reference in text []
  \gdef\refis##1{\indent\hbox to 0pt{\hss##1)~}}        % Ref list numbers)
  \gdef\journal##1, ##2, ##3, 1##4##5##6{               % Journal reference
    {\sl ##1~}{\bf ##2~}(1##4##5##6) ##3}}
\def\pr{\journal Phys. Rev., }

\def\prl{\journal Phys. Rev. Lett., }
\def\prpts{\journal Phys. Rep., }
\def\np{\journal Nucl. Phys., }
\def\pl{\journal Phys. Lett., }

\def\endreferences{\body}
\def\endpage                    %  Eject a page
  {\vfill\eject}
\def\endpaper                   %  Ways to say goodbye
  {\endmode\vfill\supereject}
\def\endit
  {\endpaper\end}
%%      Various user definitions
\def\ref#1{Ref. #1}                     %       for inline references
\def\Ref#1{Ref. #1}                     %       ditto
\def\begintable{\offinterlineskip\hrule}

\def\tablerule{\tablespace\noalign{\hrule}\tablespace}
\def\endtable{\hrule}

\def\m@th{\mathsurround=0pt }
\font\twelvesc=cmcsc10 scaled 1200
\def\cite#1{{#1}}
\def\(#1){(\call{#1})}
\def\call#1{{#1}}
\def\taghead#1{}
\def\leaderfill{\leaders\hbox to 1em{\hss.\hss}\hfill}
\def\twiddle{\lower.9ex\rlap{$\kern-.1em\scriptstyle\sim$}}
\def\bigtwiddle{\lower1.ex\rlap{$\sim$}}
\def\gtwid{\mathrel{\raise.3ex\hbox{$>$\kern-.75em\lower1ex\hbox{$\sim$}}}}
\def\ltwid{\mathrel{\raise.3ex\hbox{$<$\kern-.75em\lower1ex\hbox{$\sim$}}}}
\def\square{\kern1pt\vbox{\hrule height 1.2pt\hbox{\vrule width 1.2pt\hskip 3pt
   \vbox{\vskip 6pt}\hskip 3pt\vrule width 0.6pt}\hrule height 0.6pt}\kern1pt}
%%		EQNORDER.TEX			11/05/85	Doug E.
\catcode`@=11
\newcount\tagnumber\tagnumber=0

\immediate\newwrite\eqnfile
\newif\if@qnfile\@qnfilefalse
\def\write@qn#1{}
\def\writenew@qn#1{}
\def\w@rnwrite#1{\write@qn{#1}\message{#1}}
\def\@rrwrite#1{\write@qn{#1}\errmessage{#1}}

\def\taghead#1{\gdef\t@ghead{#1}\global\tagnumber=0}
\def\t@ghead{}

\expandafter\def\csname @qnnum-3\endcsname
  {{\t@ghead\advance\tagnumber by -3\relax\number\tagnumber}}
\expandafter\def\csname @qnnum-2\endcsname
  {{\t@ghead\advance\tagnumber by -2\relax\number\tagnumber}}
\expandafter\def\csname @qnnum-1\endcsname
  {{\t@ghead\advance\tagnumber by -1\relax\number\tagnumber}}
\expandafter\def\csname @qnnum0\endcsname
  {\t@ghead\number\tagnumber}
\expandafter\def\csname @qnnum+1\endcsname
  {{\t@ghead\advance\tagnumber by 1\relax\number\tagnumber}}
\expandafter\def\csname @qnnum+2\endcsname
  {{\t@ghead\advance\tagnumber by 2\relax\number\tagnumber}}
\expandafter\def\csname @qnnum+3\endcsname
  {{\t@ghead\advance\tagnumber by 3\relax\number\tagnumber}}

\def\equationfile{%
  \@qnfiletrue\immediate\openout\eqnfile=\jobname.eqn%
  \def\write@qn##1{\if@qnfile\immediate\write\eqnfile{##1}\fi}
  \def\writenew@qn##1{\if@qnfile\immediate\write\eqnfile
    {\noexpand\tag{##1} = (\t@ghead\number\tagnumber)}\fi}
}
\def\callall#1{\xdef#1##1{#1{\noexpand\call{##1}}}}
\def\call#1{\each@rg\callr@nge{#1}}
\def\each@rg#1#2{{\let\thecsname=#1\expandafter\first@rg#2,\end,}}
\def\first@rg#1,{\thecsname{#1}\apply@rg}
\def\apply@rg#1,{\ifx\end#1\let\next=\relax%
\else,\thecsname{#1}\let\next=\apply@rg\fi\next}
\def\callr@nge#1{\calldor@nge#1-\end-}
\def\callr@ngeat#1\end-{#1}
\def\calldor@nge#1-#2-{\ifx\end#2\@qneatspace#1 %
  \else\calll@@p{#1}{#2}\callr@ngeat\fi}
\def\calll@@p#1#2{\ifnum#1>#2{\@rrwrite{Equation range #1-#2\space is bad.}
\errhelp{If you call a series of equations by the notation M-N, then M and
N must be integers, and N must be greater than or equal to M.}}\else%
 {\count0=#1\count1=#2\advance\count1
by1\relax\expandafter\@qncall\the\count0,%
  \loop\advance\count0 by1\relax%
    \ifnum\count0<\count1,\expandafter\@qncall\the\count0,%
  \repeat}\fi}

\def\@qneatspace#1#2 {\@qncall#1#2,}
\def\@qncall#1,{\ifunc@lled{#1}{\def\next{#1}\ifx\next\empty\else
  \w@rnwrite{Equation number \noexpand\(>>#1<<) has not been defined yet.}
  >>#1<<\fi}\else\csname @qnnum#1\endcsname\fi}
\let\eqnono=\eqno
\def\eqno(#1){\tag#1}
\def\tag#1$${\eqnono(\displayt@g#1 )$$}
\def\aligntag#1\endaligntag
  $${\gdef\tag##1\\{&(##1 )\cr}\eqalignno{#1\\}$$
  \gdef\tag##1$${\eqnono(\displayt@g##1 )$$}}

\def\eqalignno#1{\displ@y \tabskip\centering
  \halign to\displaywidth{\hfil$\displaystyle{##}$\tabskip\z@skip
    &$\displaystyle{{}##}$\hfil\tabskip\centering
    &\llap{$\displayt@gpar##$}\tabskip\z@skip\crcr
    #1\crcr}}

\def\displayt@gpar(#1){(\displayt@g#1 )}

\def\displayt@g#1 {\rm\ifunc@lled{#1}\global\advance\tagnumber by1
        {\def\next{#1}\ifx\next\empty\else\expandafter
        \xdef\csname @qnnum#1\endcsname{\t@ghead\number\tagnumber}\fi}%
  \writenew@qn{#1}\t@ghead\number\tagnumber\else
        {\edef\next{\t@ghead\number\tagnumber}%
        \expandafter\ifx\csname @qnnum#1\endcsname\next\else
        \w@rnwrite{Equation \noexpand\tag{#1} is a duplicate number.}\fi}%
  \csname @qnnum#1\endcsname\fi}

\def\ifunc@lled#1{\expandafter\ifx\csname @qnnum#1\endcsname\relax}

\let\@qnend=\end\gdef\end{\if@qnfile
\immediate\write16{Equation numbers written on []\jobname.EQN.}\fi\@qnend}

\catcode`@=12
%%%%%%%%%%%%%%%%%%%%             REFORDER.TEX              %%%%%%%%%%%%%%%%%%%%
\refstyleprnp
\catcode`@=11
\newcount\r@fcount \r@fcount=0
\def\refreset{\global\r@fcount=0}
\newcount\r@fcurr
\immediate\newwrite\reffile
\newif\ifr@ffile\r@ffilefalse
\def\w@rnwrite#1{\ifr@ffile\immediate\write\reffile{#1}\fi\message{#1}}

\def\writer@f#1>>{}
\def\referencefile{%			  Stuff to write .REF file
  \r@ffiletrue\immediate\openout\reffile=\jobname.ref%
  \def\writer@f##1>>{\ifr@ffile\immediate\write\reffile%
    {\noexpand\refis{##1} = \csname r@fnum##1\endcsname = %
     \expandafter\expandafter\expandafter\strip@t\expandafter%
     \meaning\csname r@ftext\csname r@fnum##1\endcsname\endcsname}\fi}%
  \def\strip@t##1>>{}}

\def\citeall#1{\xdef#1##1{#1{\noexpand\cite{##1}}}}
\def\cite#1{\each@rg\citer@nge{#1}}	% Variable No. of args, separated by ","

\def\each@rg#1#2{{\let\thecsname=#1\expandafter\first@rg#2,\end,}}
\def\first@rg#1,{\thecsname{#1}\apply@rg}	% each@ag is a general purpose
\def\apply@rg#1,{\ifx\end#1\let\next=\relax%	  variable no. of arg. macro.
\else,\thecsname{#1}\let\next=\apply@rg\fi\next}% args separated by commas

\def\citer@nge#1{\citedor@nge#1-\end-}	% Check for M-N range (M and N numbers)
\def\citer@ngeat#1\end-{#1}
\def\citedor@nge#1-#2-{\ifx\end#2\r@featspace#1 % Single argument
  \else\citel@@p{#1}{#2}\citer@ngeat\fi}	% M-N range of arguments
\def\citel@@p#1#2{\ifnum#1>#2{\errmessage{Reference range #1-#2\space is bad.}%
    \errhelp{If you cite a series of references by the notation M-N, then M and
    N must be integers, and N must be greater than or equal to M.}}\else%
 {\count0=#1\count1=#2\advance\count1
by1\relax\expandafter\r@fcite\the\count0,%
  \loop\advance\count0 by1\relax%	  Loop from M to N
    \ifnum\count0<\count1,\expandafter\r@fcite\the\count0,%
  \repeat}\fi}

\def\r@featspace#1#2 {\r@fcite#1#2,}	% Eat spaces at beginning or end of arg
\def\r@fcite#1,{\ifuncit@d{#1}%		  Cite individual reference
    \newr@f{#1}%
    \expandafter\gdef\csname r@ftext\number\r@fcount\endcsname%
                     {\message{Reference #1 to be supplied.}%
                      \writer@f#1>>#1 to be supplied.\par}%
 \fi%
 \csname r@fnum#1\endcsname}
\def\ifuncit@d#1{\expandafter\ifx\csname r@fnum#1\endcsname\relax}%
\def\newr@f#1{\global\advance\r@fcount by1%
    \expandafter\xdef\csname r@fnum#1\endcsname{\number\r@fcount}}

\let\r@fis=\refis			% Save old \refis, redefine
\def\refis#1#2#3\par{\ifuncit@d{#1}%      Use two params #2 #3 to strip blank
   \newr@f{#1}%
   \w@rnwrite{Reference #1=\number\r@fcount\space is not cited up to now.}\fi%
  \expandafter\gdef\csname r@ftext\csname r@fnum#1\endcsname\endcsname%
  {\writer@f#1>>#2#3\par}}

\def\ignoreuncited{%   redefine \refis if ignoring uncited references
   \def\refis##1##2##3\par{\ifuncit@d{##1}%
     \else\expandafter\gdef\csname r@ftext\csname
r@fnum##1\endcsname\endcsname%
     {\writer@f##1>>##2##3\par}\fi}}

\def\r@ferr{\endreferences\errmessage{I was expecting to see
\noexpand\endreferences before now;  I have inserted it here.}}
\let\r@ferences=\references
\def\references{\r@ferences\def\endmode{\r@ferr\par\endgroup}}

\let\endr@ferences=\endreferences
\def\endreferences{\r@fcurr=0%		  Save old \endreferences, redefine
  {\loop\ifnum\r@fcurr<\r@fcount%	  Loop over refnum and produce text
    \advance\r@fcurr by 1\relax\expandafter\r@fis\expandafter{\number\r@fcurr}%
    \csname r@ftext\number\r@fcurr\endcsname%
  \repeat}\gdef\r@ferr{}\global\r@fcount=0\endr@ferences}

\let\r@fend=\endpaper\gdef\endpaper{\ifr@ffile
\immediate\write16{Cross References written on []\jobname.REF.}\fi\r@fend}

\catcode`@=12

\citeall\refto		% These macros will generate citations
\citeall\ref		%
\citeall\Ref		%

\referencefile

\def\frac#1/#2{#1 / #2}

\def\umichadd{Randall Physics Laboratory\\University of Michigan
\\Ann Arbor MI 48109-1120}

\def\oneandfourfifthsspace{\baselineskip=\normalbaselineskip
  \multiply\baselineskip by 9 \divide\baselineskip by 5}
\def\oneandthreefifthsspace{\baselineskip=\normalbaselineskip
  \multiply\baselineskip by 8 \divide\baselineskip by 5}

\font\titlefont=cmr10 scaled\magstep3
\def\bigtitle                      %  Title on title page
  {\null\vskip 3pt plus 0.2fill
   \beginlinemode \doublespace \raggedcenter \titlefont}
\def\dnu{D_X}
\def\msQ{m^2_{\tilde Q}}
\def\msu{m^2_{\tilde u}}
\def\msd{m^2_{\tilde d}}
\def\msL{m^2_{\tilde L}}
\def\mse{m^2_{\tilde e}}

\def\mul{m_{\tilde u_L}}
\def\mdl{m_{\tilde d_L}}
\def\mur{m_{\tilde u_R}}
\def\mdr{m_{\tilde d_R}}
\def\mel{m_{\tilde e_L}}

\def\mer{m_{\tilde e_R}}

\newbox\hdbox%
\newcount\hdrows%
\newcount\multispancount%
\newcount\ncase%
\newcount\ncols% This is the number of primary text columns in the table.
\newcount\nrows%
\newcount\nspan%
\newcount\ntemp%
\newdimen\hdsize%
\newdimen\newhdsize%
\newdimen\parasize%
\newdimen\spreadwidth%
\newdimen\thicksize%
\newdimen\thinsize%
\newdimen\tablewidth%
\newif\ifcentertables%
\newif\ifendsize%
\newif\iffirstrow%
\newif\iftableinfo%
\newtoks\dbt%
\newtoks\hdtks%
\newtoks\savetks%
\newtoks\tableLETtokens%
\newtoks\tabletokens%
\newtoks\widthspec%
\tableinfotrue%
\catcode`\@=11%  Allows use of "@" in macro names, like PLAIN.TEX does.
%  Debugging aid.  Writes #1 on the
%                                    user's terminal and in the log file.
\def\tstrut{\vrule height3.1ex depth1.2ex width0pt}%
\def\and{\char`\&}%  Allows us to get an `&' in the text.  This is the
%                    same as using the PLAIN TeX macro \&.
\def\tablerule{\noalign{\hrule height\thinsize depth0pt}}%
\thicksize=1.5pt%  Default thickness for fat rules.  The user should feel
%                  free to change this to his preference.
\thinsize=0.6pt%   Default thickness for thin rules.
\def\thickrule{\noalign{\hrule height\thicksize depth0pt}}%
\def\ctr#1{\hfil\ #1\hfil}%
\tablewidth=-\maxdimen%
\spreadwidth=-\maxdimen%
\def\tabskipglue{0pt plus 1fil minus 1fil}%
\centertablestrue%
\parasize=4in%
\gdef\ARGS{########}%  Produces the correct number of #'s in the preamble
\gdef\headerARGS{####}%  Same as \ARGS, but used in \header macros.
\def\@mpersand{&}%  Allows us to get alignment tab characters later
%                   when we have made the character "&" an active macro.
{\catcode`\|=13%  Make |'s locally active.
\gdef\letbarzero{\let|0}%  Globally define a macro that allows us to
%                          keep active |'s from being expanded in edef's.
\gdef\letbartab{\def|{&&}}%
\gdef\letvbbar{\let\vb|}%
}%  End of locally active |'s.
{\catcode`\&=4%  Make these alignment tabs.
\def\ampskip{&\omit\hfil&}%  This local macro skips a vertical rule.
\catcode`\&=13%  Now make &'s into active macros.
\let&0%  This allows us to expand \ampskip in the next \xdef without
\xdef\letampskip{\def&{\ampskip}}%
\gdef\letnovbamp{\let\novb&\let\tab&}
}%  End of locally active &'s.
\def\begintable{%  Here we make |'s and &'s active characters so we can
   \begingroup%
   \catcode`\|=13\letbartab\letvbbar%
   \catcode`\&=13\letampskip\letnovbamp%
   \def\multispan##1{%  We must redefine \multispan to count the number
%                       of primary columns, not physical columns.
      \mscount##1%
      \multiply\mscount\tw@\advance\mscount\m@ne%
      \loop\ifnum\mscount>\@ne \sp@n\repeat%
   }%  End of \multispan macro.
   \def\|{%
      &\omit\widevline&%
   }%
   \ruledtable%  Now we call \ruledtable to do the real work.
}%  End of \begintable macro.
\long\def\ruledtable#1\endtable{%
   \offinterlineskip%  Needed to make rules touch each other.
   \tabskip 0pt%  Needed for same reason as \offinterlineskip.
   \def\widevline{\vrule width\thicksize}%  Make outer \vrule's wider.
   \def\endrow{\@mpersand\omit\hfil\crnorm\@mpersand}%
   \def\crthick{\@mpersand\crnorm\thickrule\@mpersand}%
   \def\crthickneg##1{\@mpersand\crnorm\thickrule
          \noalign{{\skip0=##1\vskip-\skip0}}\@mpersand}%
   \def\crnorule{\@mpersand\crnorm\@mpersand}%
   \def\crnoruleneg##1{\@mpersand\crnorm
          \noalign{{\skip0=##1\vskip-\skip0}}\@mpersand}%
   \let\nr=\crnorule%  A shorter abbreviation.
   \def\endtable{\@mpersand\crnorm\thickrule}%
   \let\crnorm=\cr%  Allows us to use \cr for our own purposes.
   \edef\cr{\@mpersand\crnorm\tablerule\@mpersand}%
   \def\crneg##1{\@mpersand\crnorm\tablerule
          \noalign{{\skip0=##1\vskip-\skip0}}\@mpersand}%
   \let\ctneg=\crthickneg
   \let\nrneg=\crnoruleneg
   \the\tableLETtokens%  Get the user's extra \let's, if any.
   \tabletokens={&#1}%  We add an extra alignment tab to the beginning
   \countROWS\tabletokens\into\nrows%
   \countCOLS\tabletokens\into\ncols%
   \advance\ncols by -1%
   \divide\ncols by 2%
   \advance\nrows by 1%
   \ifcentertables
      \ifhmode \par\fi%  Make sure we are in vertical mode.
      \line{%  The final table comes out as an \hbox of width the \hsize.
      \hss%  The final table will be centered left-to-right.
   \else %
      \hbox{%
   \fi
      \vbox{%
         \makePREAMBLE{\the\ncols}%  Generate the preamble.
         \edef\next{\preamble}%  This line and the next line force the
         \let\preamble=\next%    expansion of all \ARGS tokens into the
%                                appropriate number of #'s.
         \makeTABLE{\preamble}{\tabletokens}%  Go do the \halign here.
      }%  End of \vbox.
      \ifcentertables \hss}\else }\fi%  Finish the centering effect.
%                                       It is important that no spaces
%                                       follow the two `}' here.
%  }%  End of \line.
   \endgroup%  Return all local macros and parameters to their outside
%              values.
   \tablewidth=-\maxdimen%  Reset \tablewidth to normal.
   \spreadwidth=-\maxdimen% Same for \spreadwidth.
}%  End of macro \ruledtable.
\def\makeTABLE#1#2{%  Does an \halign for the \ruledtable macro.
   {%  Start of local parameter values.
   \let\ifmath0%     These macros would cause trouble if they were to be
   \let\header0%     expanded in the following \xdef; we \let them be
   \let\multispan0%  equal to a digit, because digits can't be expanded.
   \ncase=0%
   \ifdim\tablewidth>-\maxdimen \ncase=1\fi%
   \ifdim\spreadwidth>-\maxdimen \ncase=2\fi%
   \relax%  This \relax is absolutely necessary, without it the following
   \ifcase\ncase %
      \widthspec={}%
   \or %
      \widthspec=\expandafter{\expandafter t\expandafter o%
                 \the\tablewidth}%
   \else %
      \widthspec=\expandafter{\expandafter s\expandafter p\expandafter r%
                 \expandafter e\expandafter a\expandafter d%
                 \the\spreadwidth}%
   \fi %
   \xdef\next{%  We must force the preamble to be expanded BEFORE the
      \halign\the\widthspec{%
      #1%  This is the preamble text.
      \noalign{\hrule height\thicksize depth0pt}%  Makes the top \hrule.
      \the#2\endtable%  This is the main body.
      }%  End of \halign.
   }%  End of \next.
   }%  End of local values.
   \next%  This \next must be outside of the local values, because now
}%  End of macro \makeTABLE.
\def\makePREAMBLE#1{%  This macro generates the necessary preamble for a
   \ncols=#1%  Get the number of columns desired.
   \begingroup%  Start local parameter definitions.
   \let\ARGS=0%  This is the key to the whole thing; it prevents \ARGS
%                from being expanded in the following \edef's.
   \edef\xtp{\widevline\ARGS\tabskip\tabskipglue%
   &\ctr{\ARGS}\tstrut}%  A 1-column preamble.  Gets the sizing right.
   \advance\ncols by -1%  One column has been generated; decrement the
%                         counter.
   \loop%  Append as many further columns as needed to the preamble.
      \ifnum\ncols>0 %
      \advance\ncols by -1%
      \edef\xtp{\xtp&\vrule width\thinsize\ARGS&\ctr{\ARGS}}%
   \repeat
   \xdef\preamble{\xtp&\widevline\ARGS\tabskip0pt%
   \crnorm}%  Adds the last \vrule.
   \endgroup%  End of local parameters.
}%  End of macro \makePREAMBLE.
\def\countROWS#1\into#2{%  This counts the number of rows in #1 by
   \let\countREGISTER=#2%
   \countREGISTER=0%
%  \out{In countROWS:  tokens are [\the#1]}%
   \expandafter\ROWcount\the#1\endcount%
}%
\def\ROWcount{%
   \afterassignment\subROWcount\let\next= %
}%
\def\subROWcount{%
%  \out{In subROWcount:  next is [\meaning\next]}%  Debugging aid.
   \ifx\next\endcount %
      \let\next=\relax%
   \else%
      \ncase=0%
      \ifx\next\cr %
         \global\advance\countREGISTER by 1%
         \ncase=0%
      \fi%
      \ifx\next\endrow %
         \global\advance\countREGISTER by 1%
         \ncase=0%
      \fi%
      \ifx\next\crthick %
         \global\advance\countREGISTER by 1%
         \ncase=0%
      \fi%
      \ifx\next\crnorule %
         \global\advance\countREGISTER by 1%
         \ncase=0%
      \fi%
      \ifx\next\crthickneg %
         \global\advance\countREGISTER by 1%
         \ncase=0%
      \fi%
      \ifx\next\crnoruleneg %
         \global\advance\countREGISTER by 1%
         \ncase=0%
      \fi%
      \ifx\next\crneg %
         \global\advance\countREGISTER by 1%
         \ncase=0%
      \fi%
      \ifx\next\header %
         \ncase=1%
      \fi%
      \relax%
      \ifcase\ncase %
         \let\next\ROWcount%
      \or %
         \let\next\argROWskip%
      \else %
      \fi%
   \fi%
   \next%
}%  End of macro \subROWcount.
\def\counthdROWS#1\into#2{%
\dvr{10}%
   \let\countREGISTER=#2%
   \countREGISTER=0%
\dvr{11}%
\dvr{13}%
   \expandafter\hdROWcount\the#1\endcount%
\dvr{12}%
}%
\def\hdROWcount{%
   \afterassignment\subhdROWcount\let\next= %
}%
\def\subhdROWcount{%
   \ifx\next\endcount %
      \let\next=\relax%
   \else%
      \ncase=0%
      \ifx\next\cr %
         \global\advance\countREGISTER by 1%
         \ncase=0%
      \fi%
      \ifx\next\endrow %
         \global\advance\countREGISTER by 1%
         \ncase=0%
      \fi%
      \ifx\next\crthick %
         \global\advance\countREGISTER by 1%
         \ncase=0%
      \fi%
      \ifx\next\crnorule %
         \global\advance\countREGISTER by 1%
         \ncase=0%
      \fi%
      \ifx\next\header %
         \ncase=1%
      \fi%
\relax%
      \ifcase\ncase %
         \let\next\hdROWcount%
%\out{subhdROWcount---> ncase=\the\ncase}%
      \or%
         \let\next\arghdROWskip%
      \else %
      \fi%
   \fi%
   \next%
}%
{\catcode`\|=13\letbartab
\gdef\countCOLS#1\into#2{%
   \let\countREGISTER=#2%
   \global\countREGISTER=0%
   \global\multispancount=0%
   \global\firstrowtrue
   \expandafter\COLcount\the#1\endcount%
   \global\advance\countREGISTER by 3%
   \global\advance\countREGISTER by -\multispancount
%  \out{countCOLS-->[\the\countREGISTER]}
}%
\gdef\COLcount{%
   \afterassignment\subCOLcount\let\next= %
}%
{\catcode`\&=13%
\gdef\subCOLcount{%
   \ifx\next\endcount %
      \let\next=\relax%
   \else%
      \ncase=0%
      \iffirstrow
         \ifx\next& %
            \global\advance\countREGISTER by 2%
            \ncase=0%
         \fi%
         \ifx\next\span %
            \global\advance\countREGISTER by 1%
            \ncase=0%
         \fi%
         \ifx\next| %
            \global\advance\countREGISTER by 2%
            \ncase=0%
         \fi
         \ifx\next\|
            \global\advance\countREGISTER by 2%
            \ncase=0%
         \fi
         \ifx\next\multispan
            \ncase=1%
            \global\advance\multispancount by 1%
         \fi
         \ifx\next\header
            \ncase=2%
         \fi
         \ifx\next\cr       \global\firstrowfalse \fi
         \ifx\next\endrow   \global\firstrowfalse \fi
         \ifx\next\crthick  \global\firstrowfalse \fi
         \ifx\next\crnorule \global\firstrowfalse \fi
         \ifx\next\crnoruleneg \global\firstrowfalse \fi
         \ifx\next\crthickneg  \global\firstrowfalse \fi
         \ifx\next\crneg       \global\firstrowfalse \fi
      \fi%  End of \iffirstrow.
\relax%\out{subCOL-->  ncase=[\the\ncase]}
      \ifcase\ncase %
         \let\next\COLcount%
      \or %
         \let\next\spancount%
      \or %
         \let\next\argCOLskip%
      \else %
      \fi %
   \fi%
   \next%
}%
\gdef\argROWskip#1{%
   \let\next\ROWcount \next%
}%  End of macro \argskip.
\gdef\arghdROWskip#1{%
   \let\next\ROWcount \next%
}%  End of macro \arghdROWskip.
\gdef\argCOLskip#1{%
   \let\next\COLcount \next%
}%  End of macro \argskip.
}%  End of active &'s.
}%  End of active |'s.
\def\spancount#1{%\out{spancount--->\meaning#1}
   \nspan=#1\multiply\nspan by 2\advance\nspan by -1%
   \global\advance \countREGISTER by \nspan
   \let\next\COLcount \next}%
\def\dvr#1{\relax}%
\def\header#1{%
\dvr{1}{\let\cr=\@mpersand%
\hdtks={#1}%
\counthdROWS\hdtks\into\hdrows%
\advance\hdrows by 1%
\ifnum\hdrows=0 \hdrows=1 \fi%
\dvr{5}\makehdPREAMBLE{\the\hdrows}%
\dvr{6}\getHDdimen{#1}%
{\parindent=0pt\hsize=\hdsize{\let\ifmath0%
\xdef\next{\valign{\headerpreamble #1\crnorm}}}\dvr{7}\next\dvr{8}%
}%
}\dvr{2}}%  End of macro \header.
\def\makehdPREAMBLE#1{%This macro generates the necessary preamble for a
\dvr{3}%
\hdrows=#1%  Get the number of columns desired.
{%  Start local parameter definitions.
\let\headerARGS=0%
\let\cr=\crnorm%
\edef\xtp{\vfil\hfil\hbox{\headerARGS}\hfil\vfil}%
\advance\hdrows by -1%  One row has been generated; decrement the
\loop%  Append as many further rows as needed to the preamble.
\ifnum\hdrows>0%
\advance\hdrows by -1%
\edef\xtp{\xtp&\vfil\hfil\hbox{\headerARGS}\hfil\vfil}%
\repeat%
\xdef\headerpreamble{\xtp\crcr}%
}%  End of local parameters.
\dvr{4}}%  End of \makehdPREAMBLE.
\def\getHDdimen#1{%
\hdsize=0pt%
\getsize#1\cr\end\cr%
}%  End of macro getHDdimen.
\def\getsize#1\cr{%
\endsizefalse\savetks={#1}%
\expandafter\lookend\the\savetks\cr%
\relax \ifendsize \let\next\relax \else%
\setbox\hdbox=\hbox{#1}\newhdsize=1.0\wd\hdbox%
\ifdim\newhdsize>\hdsize \hdsize=\newhdsize \fi%
\let\next\getsize \fi%
\next%
}%
\def\lookend{\afterassignment\sublookend\let\looknext= }%
\def\sublookend{\relax%
\ifx\looknext\cr %
\let\looknext\relax \else %
   \relax
   \ifx\looknext\end \global\endsizetrue \fi%
   \let\looknext=\lookend%
    \fi \looknext%
}%
\def\tablelet#1{%
   \tableLETtokens=\expandafter{\the\tableLETtokens #1}%
}%
\catcode`\@=12%  Change @'s back to their normal category code.
\hyphenation{Nil-les}

%% SPACING
%%%%%%% for preprint and hep-ph:
\oneandthreefifthsspace
\preprintno{UM-TH-95-08}
\preprintno{March 1995}
\preprintno{(revised April 1995)}
\preprintno{hep-ph/9503445}
\bigtitle{Low-energy supersymmetry with D-term
contributions to scalar masses}
\bigskip
\author
Chris Kolda$^\dagger$ and Stephen P.~Martin
\affil\umichadd
\body
\footnote{}{$^\dagger$ Address after Sept.~1, 1995:
School of Natural Sciences, Institute for Advanced Study,
Princeton, NJ 08540}

\abstract
We investigate how the predictions of the Minimal Supersymmetric Standard
Model are modified by D-term contributions to soft scalar masses, which arise
whenever the rank of the gauge group at very high energies is greater than
four. We give a parameterization of the most general such contributions that
can occur when the unbroken gauge symmetry is an arbitrary subgroup of $E_6$,
and show how the D-term contributions leave their imprint on physics at
ordinary energies.  We impose experimental constraints on the resulting
parameter space and discuss some features of the resulting  supersymmetric
spectrum which differ from the predictions obtained with universal boundary
conditions on scalar masses near the Planck scale. These include relations
between squark and slepton masses; the behavior of $\sin^2 (\beta-\alpha)$
(which determines the production cross-section for the lightest Higgs scalar
boson at an $e^+e^-$ collider) and the mass of the pseudoscalar Higgs bosons;
$R_b$ [the ratio $\Gamma(Z\rightarrow b\overline b)/\Gamma(Z\rightarrow
{\rm hadrons})$]; and mass differences between charginos and neutralinos.

\endtitlepage
\oneandfourfifthsspace

\subhead{1. Introduction}
\taghead{1.}

The mass of the Higgs scalar boson in the standard model is subject to
quadratically divergent radiative corrections of order $\delta m^2_H
\sim \Lambda^2$, where $\Lambda$ is an ultraviolet cutoff mass. This
poses a naturalness problem for $\Lambda$ much larger than the electroweak
scale. Low-energy supersymmetry (SUSY) [\cite{reviews}] presents a beautiful
solution to this problem, because all quadratic divergences cancel
order-by-order in perturbation theory. The cancellation still works if
SUSY is softly broken by scalar mass terms, scalar trilinear couplings,
and gaugino mass terms. These are exactly the parameters which determine
the masses of the superpartners of standard model particles; therefore,
in the minimal supersymmetric standard model (MSSM),
the superpartners must have
masses which do not greatly exceed the TeV range in order to maintain
the stability of the electroweak scale. Moreover, we already
know that the soft SUSY-breaking terms in the MSSM are far from arbitrary
in their flavor and phase structure. Otherwise, large flavor changing neutral
currents could be expected to manifest themselves in processes such as
$K-\overline K$ mixing, $b\rightarrow s \gamma$, and $\mu \rightarrow e
\gamma$. Arbitrary complex phases would give rise to a CP-violating
electric dipole moment for the neutron in violation of experimental bounds.
Thus there is strong circumstantial evidence in favor of some organizing
principle governing the soft SUSY-breaking terms.

Supergravity models [\cite{supergravity}] do provide just such an organizing
principle for the soft terms,  if one makes the assumption that gravity
is flavor blind. SUSY is presumed to be broken at a scale
$M_I \sim 10^{10} $ GeV in a ``hidden" sector of particles which have only
gravitational interactions with the ``visible" sector of particles familiar to
us. The SUSY breaking is transmitted to the visible sector by gravitational
effects, so that the soft SUSY-breaking terms in the visible sector are
characterized by a mass scale $m_{\rm SUSY} = M_I^2/M_{\rm Planck} \sim M_Z$
and are universal, independent of the unknown physics of the hidden sector.
This mechanism implies a simple form for the soft breaking Lagrangian  at the
Planck scale. The following terms arise:
{\phantom{\cite{sw}\cite{il}\cite{drees}\cite{hk}\cite{fhkn}\cite{llm}
\cite{kt}\cite{kmy}\cite{rsh}\cite{mn} }}

\noindent $\bullet$ A common (mass)$^2$ (denoted $m_0^2$) for all scalars in
the theory. {\phantom{\cite{op} \cite{pp} \cite{ch} \cite{dp}} }

\noindent $\bullet$ Scalar trilinear couplings which are given by the
corresponding Yukawa couplings in the superpotential multiplied
by a common mass parameter $A_0$.

\noindent $\bullet$ Scalar (mass)$^2$ terms given by the corresponding
mass in the superpotential multiplied by a common mass parameter $B_0$.

\noindent $\bullet$ A common mass $m_{1/2}$ for the gauginos.

\noindent The mass parameters $m_0$, $A_0$, $B_0$, and $m_{1/2}$ are all of
order $m_{\rm SUSY}$.
The assumption of universality may have to
be modified in non-minimal supergravity models [\cite{sw}] or in
superstring models [\cite{il}]; it is clear in any case that while they
are sufficient to avoid flavor-changing neutral currents, they are certainly
not necessary. The origins and consequences of scalar mass non-universalities
which are safe for flavor-changing neutral currents have been
discussed for example in [\cite{drees}-\cite{dp}].
{\phantom {\cite{rr}\cite{dn}\cite{cpr}\cite{an}\cite{bbo}\cite{kkrw}
\cite{copw}\cite{fhkn}\cite{op1}}}

In the MSSM, the superpotential is given by
$$
W = u { Y}_u Q H_u + d { Y}_d Q H_d + e { Y}_e L H_d +
\mu H_u H_d
$$
where $Q,L$ ($u,d,e$) are chiral superfields for the $SU(2)_L$
doublet (singlet) quarks and leptons; $H_u,H_d$ are the Higgs
doublet chiral superfields with weak hypercharge $+1/2,-1/2$; and
${Y}_u, {Y}_d, {Y}_e$ are $3\times 3$ Yukawa matrices.
The simple supergravity assumptions above tell us that the
soft SUSY-breaking Lagrangian at $M_{\rm Planck}$  is
$$
\eqalign{
L_{\rm soft} = & - m_0^2 \left [ |H_u|^2 + |H_d|^2 + \sum_{\rm families}
( |Q|^2 + |u|^2 + |d|^2 + |L|^2 + |e|^2 ) \right ]
\cr
& - A_0 (u {Y}_u Q H_u + d {Y}_d Q H_d + e { Y}_e L H_d )
+ {\rm H.c.} \cr
& - B_0 \mu H_u H_d + {\rm H.c.} \cr
& - {1\over 2} m_{1/2} (\tilde g \tilde g + \tilde W \tilde W +
\tilde B \tilde B ) + {\rm H.c.} \cr
} \eqno(universal)
$$
(Here we use the same symbol for the scalar fields as for the chiral
superfields, and $\tilde g, \tilde W, \tilde B$ are the gauginos
for $SU(3)_C, SU(2)_L, U(1)_Y$ respectively.)
This parametrization serves as a set of boundary conditions on the model.
The couplings so specified must be run down to ordinary energies
using the  renormalization group (RG) equations. This has been done by
many groups; recent examples include [\cite{rr}-\cite{op1}].

Of course, it is quite unlikely that there is no new physics between
the TeV scale and the Planck scale. A much-heralded hint of this is
the apparent unification of gauge
couplings at $M_U \approx 2 \times 10^{16}$
GeV with SUSY thresholds near the electroweak scale[\cite{unification}].
This suggests a SUSY grand unified theory (GUT) or superstring model, both
of which in turn
imply a variety of new chiral superfields and new gauge interactions
at least two  orders of magnitude below the Planck scale at which the
boundary conditions \(universal) should be applied. The existence of
this physics should be taken into account by running the gauge couplings
of the theory from $M_{\rm Planck}$ down to the weak scale, integrating
out non-MSSM fields along the way as they acquire masses.
Unfortunately, it is very difficult to anticipate with any confidence
the nature of the physics between $M_U$ and $M_{\rm Planck}$.
 For this reason, most studies
have used the approximation of applying the boundary conditions
\(universal) at $M_U$ rather than $M_{\rm Planck}$. Some recent
papers[\cite{kmy},\cite{mn},\cite{pp}]  have taken into account the
effects of RG running between $M_U$ and $M_{\rm Planck}$, which can be large
e.g.~in certain GUT models.

In this paper we will be mainly concerned with the D-term contributions
[\cite{drees}] to soft scalar masses
which should arise in any model with a
gauge group of rank $>4$ which breaks down to
$SU(3)_C \times SU(2)_L \times U(1)_Y$. Some of the phenomenological
implications of D-term contributions have recently been explored in
[\cite{hk},\cite{fhkn},\cite{kt},\cite{kmy},\cite{rsh},\cite{ch}].
We will consider here the case that the
underlying gauge group is an arbitrary rank 5 or 6 subgroup of $E_6$.
This encompasses a very wide range of special cases, including $SO(10)$
SUSY GUTs and string-inspired models including e.g. ``flipped"
$SU(5)\times U(1)$ and $SU(3)_C \times SU(3)_L \times SU(3)_R$.
The  breaking of the additional
$U(1)$ factors embedded within such groups leads to specific patterns in
the soft scalar masses which may be thought of as non-universal corrections
to the universal boundary conditions \(universal), even though they typically
arise at a much lower scale. These non-universal corrections to the
scalar masses do not imply additional contributions to
flavor-changing neutral current processes, since the contributions are the
same for each set of squarks and sleptons with
the same $SU(3)_C \times SU(2)_L \times U(1)_Y$ quantum numbers, e.g.
$(\tilde d_L, \tilde s_L, \tilde b_L)$.
The D-terms allow one to reach  certain
regions of parameter space which cannot be accessed using the universal
boundary conditions, sometimes with interesting implications for phenomenology
below the TeV scale. For example, much smaller values of $\mu$ can be obtained
for fixed values of the other parameters if the D-term contributions have
the appropriate sign and magnitude. The
Higgs pseudoscalar mass might be lower than predicted in the universal case,
and the lightest neutral Higgs eigenstate may have couplings which differ
significantly from those of a standard model Higgs scalar of the same mass.
The D-terms may also leave a dramatic
imprint on sum rule relations among the squark and slepton masses. Such
effects may eventually help to distinguish between various scenarios
for gauge symmetry breaking at very high energies.

This paper is organized as follows. In section 2 we will review the
origin of D-term contributions to soft scalar masses, and introduce a
parametrization of the most general such contributions arising in the
MSSM from arbitrary subgroups of $E_6$. Using the RG equations we will show
how the D-term contributions at very high energies leave their
imprint on TeV scale physics.
In section 3 we will discuss some of the salient phenomenological consequences
of the D-term-induced non-universality, taking into account direct and
indirect experimental constraints. In section 4 we will make some concluding
remarks.

\subhead{2. D-term contributions to soft scalar masses}
\taghead{2.}

In general, D-term contributions to the scalar masses will arise whenever
a gauge symmetry is spontaneously
broken with reduction of rank [\cite{drees}]. To understand this,
consider a toy model with a gauge group containing a product of abelian
factors $\prod_I U(1)_I$.  We take the $U(1)_I$ gauge couplings
and charges to be normalized so that the $U(1)_I$ gauge supermultiplets have
canonical kinetic terms. This determines, up to orthogonal rotations
on the index $I$, a preferred basis for the $U(1)_I$ charges which is
{\it not} invariant under general invertible
linear transformations. The gauge group is to be spontaneously broken by VEVs
for the scalar components of chiral superfields
$\Phi$ and $ \overline \Phi$ with charges $Q_{I\Phi}$ and $-Q_{I\Phi}$
respectively under $U(1)_I$. For illustrative purposes, we may for example
suppose that the spontaneous symmetry breaking is accomplished by taking
the superpotential to be
$$
W = {1\over n M^{2n-3}} (\Phi \overline \Phi)^n
\qquad\qquad (n \geq 2)
$$
with $M$ a mass parameter of order the Planck or string scale.
The supersymmetric part of the scalar potential including D-terms is given by
$$
\eqalign{
V_{\rm SUSY} = &{1\over M^{4n-6}} (|\Phi|^2 +
|\overline \Phi|^2 ) |\Phi \overline \Phi|^{2n-2} \cr
& + \sum_I {g_I^2 \over 2} \Bigl (
Q_{I\Phi}(|\Phi|^2 - |\overline \Phi|^2 )
+ \sum_a Q_{Ia} |\varphi_a|^2 \Bigr )^2
\cr }
\eqno(toyw)
$$
where the additional fields $\varphi_a$ play the role of the scalar fields
of the MSSM. In addition there are soft SUSY-breaking terms which include
$$
V_{\rm soft} = m^2 |\Phi|^2 + \overline m^2 |\overline \Phi |^2 \>.
\eqno(toysoft)
$$
The full scalar potential has a non-trivial VEV in a nearly D-flat direction:
$$
\langle \Phi \rangle^2 \approx \langle \overline \Phi \rangle^2 \approx
\left [ {- (m^2 + \overline m^2) M^{4n-6} \over (4n-2) }\right ]^{1/(2n-2)}
\eqno(toyvevs)
$$
provided that $m^2 + \overline m^2 < 0$ at the scale $\langle\Phi\rangle$.
This can be easily achieved if for example $\Phi$ has large Yukawa couplings
to some other superfields, driving the running $m^2$ negative in the
infrared. The deviation from D-flatness is given by
$$
\langle \Phi \rangle^2 - \langle \overline \Phi \rangle^2 \approx
{1\over 2}({\overline m}^2 -  m^2 )/ \sum_I g_I^2 Q_{I\Phi}^2
\> .
\eqno(toydev)
$$
After integrating out the fields $\Phi$ and  $\overline \Phi$, one obtains
corrections to the soft scalar masses of the remaining fields $\varphi_a$:
$$
\Delta m^2_a = \sum_I Q_{Ia} d_I \> ; \qquad \>\>
d_I = {1\over 2} (\overline m^2 -  m^2) g_I^2 Q_{I\Phi}
/\sum_J g_J^2 Q_{J\Phi}^2
\> .
\eqno(toydterms)
$$
{}From \(toydterms) we see that the D-term contributions do not depend on
the precise form of the superpotential \(toyw), and only depend on the
soft masses $m^2$, $\overline m^2$ and the charges under
the $U(1)_I$ gauge groups which participate in the breaking.
The VEVs \(toyvevs) reduce the rank of the gauge group by 1, with
the surviving $N-1$ $U(1)$'s having charges which are linear combinations
of the $N$ original $U(1)_I$ charges. The contributions to the (mass)$^2$
of the surviving scalar fields are just proportional to their charges under
the original $U(1)_I$ gauge groups, with proportionality constants $d_I$,
and are always of roughly the same order as the original soft scalar
(mass)$^2$ (i.e.~$m^2_{\rm SUSY}$),
even though the scale of spontaneous symmetry breaking set by \(toyvevs)
is many orders of magnitude larger [\cite{drees}].

In a general situation, the rank reduction process exemplified above
may be iterated several times; then it is clear that the total D-term
contributions to the surviving scalar fields from the spontaneously broken
$U(1)_I$ are of the form
$$
\Delta m^2_a = \sum_I Q_{Ia} D_I
\eqno(generalDs)
$$
where now the parameters $D_I$ reflect the more complicated features
of the symmetry breaking process, and in general parameterize our ignorance
of such details as which fields are getting VEVs and possible hierarchies
in the VEVs.

In addition, if the spontaneously broken gauge groups included non-abelian
generators $T^\alpha$ which commute with all of the generators of
$SU(3)_C \times SU(2)_L \times U(1)_Y$, these could contribute residual
soft mass terms of the form
$$
\Delta L_{\rm soft}  = - D_\alpha (\varphi^\dagger T^\alpha \varphi)
+ {\rm H.c.} \> .
\eqno(nonabeliands)
$$
However, in the MSSM we may dismiss this possibility if the underlying
gauge group is a subgroup of $E_6$, or more generally if it does not
mix families. This is because the only distinct chiral superfields with
the same $SU(3)_C \times SU(2)_L \times U(1)_Y$ quantum numbers in the MSSM
are $L$ and $H_d$, so that \(nonabeliands) would have to be of the form
$$
\Delta L_{\rm soft}  = - D_L L^\dagger H_d + {\rm H.c.}
\eqno(forbidden)
$$
which would necessarily imply R-parity violation. So we need only
consider $U(1)_I$ D-terms. [It might be interesting to consider
D-terms that could arise in variants of the MSSM with R-parity violation,
or with family-dependent gauge symmetries (see e.g.~[\cite{br}]).]

The group $E_6$ contains two $U(1)$ factors not contained in the standard
model gauge group $SU(3)_C \times SU(2)_L \times U(1)_Y$.
One may choose a basis for the two extra
$U(1)$ factors in an arbitrary way, but it is important to realize that
the original $U(1)_I$ groups (with canonical kinetic terms) above the
symmetry-breaking scale will in general
be linear combinations of the two chosen $U(1)$ basis groups and the
surviving weak hypercharge $U(1)_Y$. We choose as a basis
$U(1)_Y$, $U(1)_X$, and $U(1)_S$ as in Table 1.

\subhead{Table 1: $U(1)$ charges of MSSM chiral superfields}
\begintable
$~{}~$|$~Q~$|$~u~$|$~d~$|$~L~$|$~e~$|$~H_u~$|$~H_d~$
\crthick
$~U(1)_Y~$|$~{1/6}~$|$~{-2/3}~$|$~{1/3}~$|$~{-1/2}~$|$~1~$|$~{1/2}~$|$
{}~{-1/2}~$\cr
$~U(1)_X~$|$~{-1/3}~$|$~{-1/3}~$|$~{1}~$|$~{1}~$|$~{-1/3}~$|$~{2/3}~$|$
{}~{-2/3}~$\cr
$~U(1)_S~$|$~{-1/3}~$|$~{-1/3}~$|$~{-2/3}~$|$~{-2/3}~$|$~{-1/3}~
$|${2/3}$|$~{1}~$
\endtable\vskip .3cm

One of the reasons behind this choice is that within a $\bf 27$ of $E_6$ there
are two standard model singlet components $\nu$ and $S$, which carry
$[U(1)_Y,U(1)_X,U(1)_S]$ charges $[0,-5/3,0]$ and $[0,0,-5/3]$
respectively. Thus $U(1)_X$ could be broken by a VEV for the scalar
component of the field $\nu$ (which carries lepton number $-1$), and
$U(1)_S$ could be broken by the field $S$ (which carries lepton number 0).
The normalizations are chosen so that the largest
charge appearing in Table 1 is unity; in order to obtain the correct
normalization for unification into $E_6$ (or any of its subgroups),
these charges should be multiplied by $\sqrt{3\over 5}$,
$3\over 2 \sqrt{10}$, $3\over 2 \sqrt{10}$ respectively.
The gauged $B-L$ subgroup of $E_6$ is given by
$$
U(1)_{B-L} = {4\over 5} U(1)_Y - {3\over 5} U(1)_X
\> .$$

There is also a non-abelian $SU(2)$ factor within $E_6$ which commutes
with all of the surviving $SU(3)_C\times SU(2)_L \times U(1)_Y$
gauge generators. It generates simultaneous
rotations $L\leftrightarrow H_d$, $\nu \leftrightarrow S$, and
$d \leftrightarrow h$, where $h$ is a non-MSSM color anti-triplet.
As we mentioned above, the off-diagonal generators of this $SU(2)$ are
prohibited by R-parity conservation from contributing to soft-scalar
masses via D-terms. Its existence means, however, that there is a potential
ambiguity in assigning $U(1)$ charges to $H_d$, $L$, and $d$, stemming
from different embeddings of these fields into remnants of a $\bf 27$
for certain subgroups of $E_6$. The $L\leftrightarrow H_d$
ambiguity is easily fixed just by the choice of $U(1)_X$ versus
$U(1)_S$. [Note that under $U(1)_X \leftrightarrow U(1)_S$,
the charges of $L$ and $H_d$ are exchanged and those of
$(Q,u,e,H_u)$ remain invariant.]
The $U(1)$ charges of $d$ are then also uniquely fixed if we  demand
that that the MSSM Yukawa couplings are allowed by the unbroken
gauge group. Other  assignments of the $U(1)_{X,S}$ charges
for $d$ would imply that $B-L$ cannot be a gauged $U(1)$ symmetry, a not
particularly attractive possibility which we decline to pursue here.

The D-term contributions to the soft scalar masses of the MSSM
arising from spontaneous breakdown of any subgroup of $E_6$ may
therefore be parameterized by:
$$
\eqalign{
\Delta \msQ \> = \> & {1\over 6} D_Y - {1\over 3} \dnu - {1\over 3} D_S;
\cr
\Delta \msu \> = \> & -{2\over 3} D_Y - {1\over 3} \dnu - {1\over 3} D_S;
\cr
\Delta \msd \> = \> & {1\over 3} D_Y + \> \dnu - {2\over 3} D_S; \cr
\Delta \msL \> = \> & -{1\over 2} D_Y +  \> \dnu - {2\over 3} D_S ;
\cr}\qquad\qquad
\eqalign{
\Delta \mse \> = \> & \> D_Y - {1\over 3} \dnu - {1\over 3} D_S; \cr
\Delta m_{H_u}^2 \> = \> & {1\over 2} D_Y
+ {2\over 3} \dnu +{2\over 3} D_S; \cr
\Delta m_{H_d}^2 \> = \> & -{1\over 2} D_Y - {2\over 3}\dnu +  D_S \> .\cr
}\eqno(mssmds)
$$
These corrections to the scalar masses should be applied at the scale
$M_D$ of spontaneous symmetry breaking.
In order to see how these contributions affect physics at ordinary energies,
we must use the RG equations. For the squarks and sleptons of the
third family and the Higgs scalars, these are given by
$$
\eqalign{
16 \pi^2 {d \msQ \over dt} &=
-{32\over 3} g_3^2 M_3^2 - 6 g_2^2 M_2^2 - {2\over 15} g_1^2 M_1^2
+ {1\over 5} g_1^2 S + 2 y_t^2 \Sigma_t^2 +
2 y_b^2 \Sigma_b^2 \cr
16 \pi^2 {d \msu \over dt} &=
-{32\over 3} g_3^2 M_3^2 - {32\over 15} g_1^2 M_1^2
- {4\over 5} g_1^2 S + 4 y_t^2 \Sigma_t^2 \cr
16 \pi^2 {d \msd \over dt} &=
-{32\over 3} g_3^2 M_3^2 - {8\over 15} g_1^2 M_1^2
+ {2\over 5} g_1^2 S + 4 y_b^2 \Sigma_b^2 \cr
16 \pi^2 {d \msL \over dt} &=
- 6 g_2^2 M_2^2 - {6\over 5} g_1^2 M_1^2
- {3\over 5} g_1^2 S + 2 y_\tau^2 \Sigma_\tau^2 \cr
16 \pi^2 {d \mse \over dt} &=
- {24\over 5} g_1^2 M_1^2
+ {6\over 5} g_1^2 S + 4 y_\tau^2 \Sigma_\tau^2 \cr
16 \pi^2 {d m^2_{H_u} \over dt} &=
- 6 g_2^2 M_2^2 - {6\over 5} g_1^2 M_1^2
+ {3\over 5} g_1^2 S + 6 y_t^2 \Sigma_t^2 \cr
16 \pi^2 {d m^2_{H_d} \over dt} &=
- 6 g_2^2 M_2^2 - {6\over 5} g_1^2 M_1^2
- {3\over 5} g_1^2 S
+ 6 y_b^2 \Sigma_b^2 + 2 y_\tau^2 \Sigma_\tau^2 \cr
}$$
where $t \equiv {\rm ln}(Q/Q_0)$; $M_{3,2,1}$ are the running gaugino masses;
$y_{t,b,\tau}$ are the running Yukawa couplings of the third family;
$A_{t,b,\tau}$ are the corresponding scalar trilinear couplings;
$\Sigma_t^2 \equiv m^2_{H_u} + m^2_{\tilde t_L} + m^2_{\tilde t_R}
+ A_t^2 $;
$\Sigma_b^2 \equiv m^2_{H_d} + m^2_{\tilde b_L} + m^2_{\tilde b_R}
+ A_b^2 $;
$\Sigma_\tau^2 \equiv m^2_{H_d} + m^2_{\tilde \tau_L}
+ m^2_{\tilde \tau_R} + A_\tau^2 $; and
$$
S \equiv {\rm Tr}(Ym^2) = m_{H_u}^2 - m_{H_d}^2 + \sum_{\rm families}
(\msQ - 2 \msu + \msd - \msL + \mse)\> .
$$
The $U(1)_Y$ gauge couplings $g_1$ and $\alpha_1$ are taken to be in
a GUT normalization throughout this paper.
The RG equations of the first and second family squarks and sleptons
are the same except that the terms proportional to Yukawa couplings do
not appear.

Now the quantities $\Sigma_t^2$, $\Sigma_b^2$ and $\Sigma_\tau^2$
are unaffected when \(mssmds) are applied, and this persists
as the scalar masses evolve according to the RG equations.
So the only changes in the
running of the soft scalar masses occur because of the presence of the
terms proportional to $S$. One finds
$$
{dS \over dt} = {66\over 5} {\alpha_1\over 4 \pi}S
$$
which has the solution
$$
S(t) = S(t_0) {\alpha_1(t) / \alpha_1(t_0)}\>.
$$
At the scale $M_D$ associated with spontaneous symmetry breaking,
we have from \(mssmds)
$$
S(t_D) = 11 D_Y + {4\over 3} \dnu -{1\over 3} D_S
$$
Then at any scale lower than $t_D$ we find that the change induced
in the soft scalar masses by virtue of \(mssmds) (compared to a
template model in which the D-terms are not present and all other
parameters are the same) is given by
$$
\Delta m_a^2(t) = Y_a \left [ r D_Y - {4 (1-r)\over 33} \dnu +
{(1-r)\over 33} D_S \right ] + Q_{X a} \dnu + Q_{Sa} D_S
\eqno(weakscaleDs)
$$
with $r(t) \equiv \alpha_1(t) / \alpha_1(t_D)$ and $Y_a$, $Q_{X a}$,
$Q_{S a}$ the charges of the scalar $\varphi_a$
under $U(1)_Y$, $U(1)_X$, $U(1)_S$ respectively.
In general $r$ decreases as one moves to lower scales;
for $M_D$ near the GUT scale, $r \approx .43$ near the electroweak scale.
So, to a good approximation, the effect of RG running of the scalar masses
is simply to reduce the $D_Y$ contribution by a factor of $r$, while
leaving the $\dnu$ and $D_S$ contributions untouched.
If one is treating $D_Y$, $\dnu$, and $D_S$ as parameters of our ignorance
regarding the rank reduction mechanism, one may as well impose them at
e.g.~the GUT scale or at the electroweak scale.
The error incurred in doing so  can always be simply
absorbed into a redefinition of $D_Y$. We will henceforth refer to
the effective value of $D_Y$ at the electroweak scale as $\widehat D_Y$;
it is equal to the quantity in square brackets in \(weakscaleDs).

In some special cases, it is possible to make statements about the
relative sizes of $D_Y$, $\dnu$, $D_S$. For example, in an $SO(10)$
SUSY GUT, in the limit that $SO(10)$ breaks to $SU(3)_C \times SU(2)_L
\times U(1)_Y$ immediately at the GUT scale, one has
$$
D_Y = D_S = 0; \qquad \dnu \not= 0 \> .
$$
In the case that $SO(10)$ breaks down to a smaller group
[e.g.~$SU(4)_C \times SU(2)_L \times U(1)_R$]
and only then to the MSSM far below the GUT scale, one has
$$
D_S=0; \qquad D_Y, \dnu \not= 0
\> .
$$
However, there is another possible embedding of the MSSM into a
$SU(4)_C \times SU(2)_L \times U(1)$ subgroup of $E_6$ which does not fit
within $SO(10)$; this would yield instead
$$
\dnu = 0; \qquad D_Y, D_S \not= 0
\> .$$
In the case that the gauge group is
a rank 5 subgroup of $E_6$ obtained from a Calabi-Yau superstring
compactification [\cite{chsw}], one has
$$
\dnu = D_S \not= 0
$$
(with $D_Y$ possibly non-zero)
because of the  mandatory [\cite{witten}] presence of the subgroup
$U(1)_{X^\prime} = U(1)_X + U(1)_S$ in such models.
It is useful to note that in general the magnitude of $D_Y$ is expected
to vanish in the limit that $U(1)_Y$ has a vanishing mixing angle
with the spontaneously broken $U(1)$ subgroups, a limit which is
often approximately realized when the scale of spontaneous symmetry
breaking is not far below the GUT scale.

Experimental limits on sfermion masses imply constraints on $D_Y$, $D_X$,
and $D_S$. The condition that the right-handed selectron mass be above
the present experimental limit amounts to approximately
$$ m_0^2 > (45 \> {\rm GeV})^2 (1 - |\cos 2 \beta |) - .15 m_{1/2}^2
- \widehat D_Y + {1\over 3} \dnu + {1\over 3} D_S \> ,
\eqno(msercons)
$$
while the lower limit on the mass of the sneutrino from its contribution
to the invisible width of the $Z$ boson is approximately given by
$$ m_0^2 > (41 \> {\rm GeV})^2 (1 + 2.5 |\cos 2 \beta |) - .5 m_{1/2}^2
+ {1\over 2} \widehat D_Y  - \dnu + {2\over 3} D_S \> .
\eqno(msnucons)
$$
(We assume $\tan\beta > 1$ throughout this paper,
so $\cos 2\beta < 0$.)
Either \(msercons) or \(msnucons) typically sets the lower limit on
$m_0^2$ for given values of $D_Y$, $D_X$, $D_S$.

Note that if $\widehat D_Y - {1\over 3} \dnu - {1\over 3} D_S $
and $- {1\over 2} \widehat D_Y  + \dnu - {2\over 3} D_S $ are both large
and positive,
one may entertain the interesting and unusual possibility that $m_0^2$
is significantly negative.  Negative running squared masses for
squarks and sleptons at high scales do not necessarily imply that the squark
and slepton fields acquire VEVs, as long as there are no exactly D-flat
and F-flat directions involving only MSSM fields
in the supersymmetric part of the scalar potential
at high energies. The correct place to study the scalar potential is at
the scale of the putative VEV, which is typically much smaller than
$M_U$ or $M_{\rm Planck}$;
at this scale contributions to the running scalar (mass)$^2$
terms from gaugino loops will push them positive even if they were negative
at very high scales, eliminating the purported VEV.
The MSSM does have many directions which are exactly
D-flat and F-flat at the renormalizable level. However, these can be lifted by
non-renormalizable terms in the superpotential and by D-terms from additional
gauge interactions which remain unbroken in the high energy theory.
If for example $U(1)_S$ survives down to an intermediate scale
$M_I \sim \sqrt{M_Z M_U}$ [as suggested by (2.3) with $n=2$], then all D-flat
directions above $M_I$ would have to involve VEVs for $H_u$ or $H_d$,
since all other chiral superfields in the MSSM have $U(1)_S$ charges of
the same (negative) sign. One can show that all such flat directions can be
lifted in models with e.g.~chiral superfields and superpotential
interactions for gauge-singlet
neutrinos. Therefore, to maintain maximal generality, positivity
constraints on linear combinations of scalar
(mass)$^2$ terms which correspond to D-flat and F-flat directions of the MSSM
at the renormalizable level should be imposed only at $M_I$, which we take to
be
$10^{10}$ GeV. Such constraints are easily incorporated into computer programs
which calculate the sparticle spectrum numerically, as described in the next
section.

Even when the D-terms vanish, there is in the MSSM a
possibility for $m_0^2$ at the unification scale (or at least its effective
value as seen from low energies) to be very slightly negative, but only if
$m_{1/2}$ is large enough.
Numerically, we find that in models which satisfy all other constraints,
$m_0^2 \ge 0$ for $m_{1/2} < 100$ GeV, and $m_0^2 > -$(50 GeV)$^2$ when
$m_{1/2} < 400 $ GeV if D-terms are negligible.
Also note that a negative value of $D_S$ pushes all squark and slepton
masses higher while pushing the Higgs (mass)$^2$ parameters lower. For
this reason, we find that $D_S$ has essentially no lower bound from
experimental constraints on sfermion masses alone, but turns out to be
constrained by Higgs mass bounds and the stability of the
Higgs scalar potential.

Each of the contributions $D_Y$, $\dnu$, $D_S$ will split $m^2_{H_u}$
and $m^2_{H_d}$ (in addition to the splitting induced by the top
Yukawa coupling) so that there is a potentially significant impact on
electroweak symmetry breaking. In particular, the value of $\mu$ needed
for correct symmetry breaking (for other parameters fixed) may be
raised or lowered. To understand this, we may consider the tree-level
relation
$$
\mu^2 =  {1\over |\cos 2 \beta |}
(m^2_{H_d} \cos^2 \beta - m^2_{H_u} \sin^2 \beta ) -
{1\over 2} M_Z^2
\> .
$$
The change in $\mu^2$ induced by the D-terms is therefore approximately
$$
\Delta (\mu^2) = {1\over |\cos 2 \beta  |}\left [
-{1\over 2} \widehat D_Y - {2\over 3} \dnu + (\cos^2 \beta - {2\over 3}
\sin^2 \beta) D_S \right ] \> .
\eqno(changemu)
$$
In models with universal boundary conditions and no D-terms, $|\mu|$
tends to scale with max$(m_0,m_{1/2})$. If the quantity in square brackets
in \(changemu) is negative, $|\mu|$ can be substantially smaller
than is otherwise allowed, with important implications for the
neutralino and chargino masses and mixing angles and for the Higgs scalar
sector.

The D-terms can also have a substantial impact on mass relations
for the squarks and sleptons. For example, a sum rule given in [\cite{mr}]
for the first and second family squarks and sleptons is modified to
$$
2 m^2_{\tilde u_R} - m^2_{\tilde d_R} - m^2_{\tilde d_L}
+ m^2_{\tilde e_L} - m^2_{\tilde e_R} +
{10\over 3} \sin^2 \theta_W M_Z^2 |\cos 2 \beta | =
-{10\over 3} \widehat D_Y
\eqno(sumrulea)
$$
Note that $\dnu$ and $D_S$ only affect this sum rule indirectly
(and weakly) through $\widehat D_Y$, while a large contribution to $D_Y$
can cause a substantial deviation. {\phantom{\cite{mr}\cite{jlc1}\cite{ff}
\cite{fpmt}}}

Another interesting sum rule is [\cite{mr}]
$$
m_{\tilde u_R}^2 - m^2_{\tilde d_R}
= (.01) M_{\tilde g}^2 - |\cos 2 \beta | (43 \>
{\rm GeV})^2 - \widehat D_Y + {1\over 3} D_S - {4\over 3} \dnu
\> .
\eqno(sumruleb)
$$
The first two terms on the RHS of \(sumruleb) are generally small and
of opposite sign,
so that $m_{\tilde u_R}$ and $m_{\tilde d_R}$
are always very close to equal in models
without D-term or other non-universality; this relationship can be modified in
a dramatic way by the D-term contributions
as we will see in the next section. Using this and similar tests, a
substantial deviation from the $D_Y = \dnu = D_S = 0$ condition might
be discernible at an $e^+e^-$ collider [\cite{jlc1}-\cite{fpmt}],
yielding a strong (if not unambiguous) clue to physics at very high energies.

Of course, D-term contributions will generally not be the only
non-universal imprints left on the soft scalar masses by physics
at very high energies. The very presence of additional gauge groups
at high energies means that there will be additional contributions
to the RG running. In addition, there will be the effects of running
the RG equations between $M_U$ and $M_{\rm Planck}$. In order
to keep our exploration of the parameter space within reasonable
limits, we will neglect such effects insofar as they cannot
be absorbed into the ignorance parameters $D_Y$, $\dnu$
and $D_S$. This presumes that the additional gauge symmetry is broken not
far below $M_U$ and that all interactions above $M_U$ are
reasonably weak so that threshold and running effects are not
overwhelming. This situation is preferred anyway if the apparent
unification of gauge couplings is assumed to be not accidental.

It may be useful, however, to introduce a parameterization
of the most general family-independent scalar mass
non-universality at $M_U$. There is of course no unique way to pick such a
parametrization, but we clearly want to arrange that the parameters $D_Y$,
$\dnu$, $D_S$ appear also in the more general parameterization,
that RG running can be formulated in as transparent a way as possible,
and that the parameterization be invertible in terms of the running
scalar squared masses at the scale where they are presumed to be
family-independent. Since there are 7 independent masses
($\msQ,\msu,\msd,\msL,\mse, m^2_{H_u}, m^2_{H_d}$), there should be
6 parameters (in addition to the common $m_0^2$) which describe
the non-universality at the input scale.
We find it convenient to choose the contributions to scalar masses from
the extra three parameters to be just proportional to those from
the $SU(3)_C$, $SU(2)_L$, $U(1)_Y$
gaugino loops to the RG equations for the scalar masses; we will call
these contributions $K_3$, $K_2$ and $K_1$ respectively. Thus one may
parameterize the most general family-independent but non-universal
scalar masses at the input scale (e.g. $M_U$ or $M_{\rm Planck}$ in a
given theory) as follows:
$$
\pmatrix{ \msQ \cr \msu
\cr\msd\cr \msL \cr \mse \cr m_{H_u}^2 \cr m_{H_d}^2 \cr}
= \pmatrix{ 1 &  \> 1/6 & -1/3 & -1/3 & 1 & 1 & 1/36 \cr
             1 &   -2/3 & -1/3 & -1/3 & 1 & 0 & 4/9 \cr
             1 & \> 1/3 &   1  & -2/3 & 1 & 0 & 1/9 \cr
             1 & -1/2   &   1  & -2/3 & 0 & 1 & 1/4 \cr
             1 &  \> 1  & -1/3 & -1/3 & 0 & 0 &  1  \cr
             1 & \> 1/2 & 2/3  &  2/3 & 0 & 1 & 1/4 \cr
             1 & -1/2   & -2/3 &   1  & 0 & 1 & 1/4 \cr }
\pmatrix{m_0^2 \cr D_Y \cr \dnu \cr D_S \cr K_3 \cr K_2 \cr K_1 \cr }
\> . \eqno(magnificent7)
$$
This matrix is invertible, so that in principle, if one were given the scalar
(mass)$^2$ parameters at the input scale, one could recover
$m_0^2$, $D_{Y,X,S}$ and $K_{3,2,1}$:
$$
\pmatrix{m_0^2 \cr D_Y \cr \dnu \cr D_S \cr K_3 \cr K_2 \cr K_1 \cr }
= \pmatrix{ -1/2 &  -2 &  5/2& 1/6 & 1/2 & -7/3 & 8/3 \cr
             3/10 & -3/5 & 3/10 & -3/10 & 3/10 & 0 & 0 \cr
             -3/10 & 3/5  & -3/10 & 1/10 & -3/10 & 1 & -4/5 \cr
             -3/10 &  3/5  & -3/10 & -1/2 & -3/10 & 1 &  -1/5  \cr
             1/2 & 5/3  & -7/6 & -1/2&-1/2 & 5/3 & -5/3 \cr
             3/4 & 3/4  &  -3/2 &1/4& -1/4 & 5/4 & -5/4 \cr
             0 & 3  & -3 &   0  & 0 & 3 & -3 \cr }
\pmatrix{ \msQ \cr
\msu \cr \msd \cr \msL \cr \mse \cr m_{H_u}^2 \cr m_{H_d}^2 \cr} \> .
$$
Of course, in practice it is difficult to imagine being able to
reconstruct the soft masses at the input scale, especially
$m_{H_u}^2$ and $m_{H_d}^2$. More importantly, the fact that
the parameterization is invertible means that it is also general.
This parameterization has several other nice features. First, the $K_i$
contributions do not affect $S$, so that the running of the
sfermion masses of the first two families are unaffected. Thus
the imprint of non-zero $K_i$ on the squark and slepton
masses of the first two families is exactly the same at the TeV
scale as at the input scale.
In particular, the sum rules \(sumrulea) and \(sumruleb) are not
affected by non-zero $K_i$s.
At the low scale, the scalar masses of the first two families are given
by an equation of the same form as \(magnificent7),
with $\dnu$ and $D_S$ unchanged, and with
the replacements $D_Y \rightarrow \widehat D_Y$ and
$K_i \rightarrow \widehat K_i$, where $\widehat D_Y$ is just as before and
$$
\eqalign{\widehat K_i &= K_i + C_i(t); \cr
C_i(t) & = \pmatrix{{3/5} \cr {3/4} \cr {4/3 }\cr }
{2\over \pi} \int_t^{t_U} dt\> \alpha_i(t) M_i(t)^2,
\cr }\eqno(khats)
$$
taking into account one-loop RG contributions
and neglecting the small Yukawa couplings.
Non-zero $K_i$s only enter the RG equations for
the third family squarks and sleptons and the Higgs scalars
through the combinations $\Sigma_t^2$, $\Sigma_b^2$ and $\Sigma_\tau^2$.
These give additional contributions which may be described
to a good approximation by the parameterization
$$\eqalign{&\Delta m^2_{\tilde t_L, \tilde b_L}  = -X_t - X_b;
\qquad \>\> \Delta m^2_{\tilde t_R} = -2 X_t;
\qquad \>\> \Delta m^2_{\tilde b_R} = -2 X_b;
\qquad \>\> \Delta m^2_{\tilde \tau_L,\tilde\nu_\tau} = -X_\tau;
\cr & \Delta m^2_{\tilde \tau_R} = -2 X_\tau;
\qquad \>\> \Delta m^2_{H_u} = -3 X_t;
\qquad \>\> \Delta m^2_{H_d} = - 3 X_b - X_\tau \> .
\cr }
$$
When $\tan\beta$ is not large, $X_b$ and $X_\tau$ are negligible.

In the MSSM, there is a large uncertainty in the numerical contribution
of gluino loops to squark masses from the RG running, because the strong
coupling constant is not very precisely known and because the function
$C_3(t)$ in \(khats) runs quickly with scale below 1 TeV. Note that this
uncertainty can effectively be absorbed into the non-universality
parameter $K_3$. Similarly, the additional RG effects from MSSM gaugino loops
(but not heavy non-MSSM gaugino loops) between $M_U$ and $M_{\rm Planck}$,
or above any intermediate scale in which the gauge couplings run differently
(due to e.g.~thresholds from heavy vector-like chiral superfields)
can always be absorbed into the $K_i$. Other new physics at very high energies
(e.g.~gaugino loops for gauge generators which do not commute with
$SU(3)_C\times SU(2)_L \times U(1)_Y$)
may influence low energy physics through more complicated
combinations of the parameters given above.

\subhead{3. Phenomenological constraints and numerical results}
\taghead{3.}

In this section we will study some of the implications of D-term
non-universality on SUSY phenomenology. To do so consistently requires that we
simultaneously impose a variety of constraints from correct electroweak
symmetry breaking, absence of color-breaking and charge-breaking global
minima of the scalar potential, direct and indirect sparticle and
Higgs boson mass limits [\cite{pdb}], and a neutralino LSP. It is convenient
to study the resulting constrained parameter space using a computer program
which generates models randomly, imposing the constraints and calculating
the resulting sparticle masses, mixing angles and couplings numerically.
Here we describe the results of such an investigation, using a
computer program similar to the one described in [\cite{kkrw}].
We first treat some special cases obtained by adding D-term contributions
to a ``template" model with universal boundary conditions, to illustrate
how experimental constraints limit the size of the D-terms and how the
SUSY phenomenology is in turn modified by the D-terms. We will then turn
to a much more general study of the parameter space allowed by arbitrary
values of all parameters as constrained by experiment. In all cases,
the D-terms are applied at a scale $M_U \approx 2\times 10^{16}$ where
the $SU(3)_C \times SU(2)_L\times U(1)_Y$ gauge couplings unify, and we
retain the universal boundary conditions on gaugino masses and scalar
trilinear couplings at $M_U$.

We choose as a template model the MSSM with the following parameters fixed:
$$
\eqalign{
&m_0^2 = (100 \> {\rm GeV})^2; \qquad\qquad m_{1/2} = A_0 = 100 \>{\rm GeV};
\cr
&m_{\rm top} = 175 \> {\rm GeV};\qquad\qquad
\tan \beta = 3; \qquad\qquad {\rm sign }(\mu)= +;
\cr
&D_Y  = \dnu = D_S = 0 \> .
\cr }
$$
Now consider turning on the D-terms. As explained in section 2, the
following (approximate) directions in parameter space may be theoretically
preferred:
$$
\eqalign{
&{\rm (A)} \>\> \dnu \not= 0; \qquad D_Y=D_S=0, \cr
&{\rm (B)} \>\> D_S \not=0; \qquad D_Y=\dnu=0, \cr
&{\rm (C)} \>\> D_S = \dnu \not=0; \qquad D_Y = 0\>. \cr
}
$$
Therefore, we will investigate as examples what happens in each of cases
A, B, and C separately.

\noindent
Case A: In Figure 1 we plot $\mel$, $\mer$, $m_h$ and $m_A$ as a function
of $\dnu$, with $D_Y=D_S=0$ and all other parameters fixed as in the
template model. (In our notation, $h$ and $A$ are the lightest Higgs
scalar and pseudoscalar boson, respectively.) The lower bound on $\dnu$
in this example is set by the sneutrino contribution to the invisible width
of the $Z$ boson, which translates to a limit $\mel > 82$ GeV in this example
because of the sum rule
$$m_{\tilde e_L}^2 - m_{\tilde \nu}^2 = | \cos 2 \beta | M_W^2\> .
\eqno(elersumrule)
$$
The upper bound on $\dnu$ is set by the experimental bound
[\cite{pdb}] $m_h > 44$ GeV. In the remaining range $-(95\>{\rm GeV})^2 <
\dnu < (160\> {\rm GeV})^2$, all other sparticles have experimentally
allowed masses. Throughout most of this allowed range, $m_A$ is sufficiently
large compared to $m_h$ so that $h$ behaves essentially like a Standard
Model Higgs boson. However, as $\dnu$ approaches its upper bound, $m_A$
and $m_h$ become small together, as does $\mu$. This has potentially
important implications for Higgs production via $e^+e^- \rightarrow Zh$,
the rate for
which is proportional to the quantity $\sin^2(\beta-\alpha)$. [In the limit
$m_A \gg m_h$, one has $\sin^2(\beta-\alpha) \approx 1$; the heavier Higgs
scalar eigenstates decouple.]
For $\dnu$ very close to its upper bound, the lightest
Higgs mass eigenstate $h$ has couplings which are not like a Standard Model
Higgs boson; for $\dnu > (156\>{\rm GeV})^2$, we find in this example
$\sin^2 (\beta - \alpha) < 0.6$. However, as long as
$\dnu < (135\>{\rm GeV})^2$, one has $\sin^2 (\beta-\alpha) > 0.9$ here.

\noindent
Case B: In Figure 2 we show $\mel$, $\mer$, $m_h$ and $m_A$ this time
as a function of $D_S$ added to the template model. As can be clearly seen
from \(mssmds), $\mel$, $\mer$ and all other squark and slepton masses
grow monotonically with increasingly negative $D_S$. Thus the lower bound on
$D_S$ is typically set by the lightest Higgs mass bound; here we find
$D_S > -(325\>{\rm GeV})^2$. Conversely, the upper bound on $D_S$ is
usually set by the lower experimental bound on either $\mer$ or (as
in this case) $m_{\tilde \nu}$. For $D_S$ near its experimental lower bound,
$h$ behaves very differently from a Standard Model Higgs, with
$\sin^2 (\beta-\alpha) < 0.6$ for $D_S < -(310\>{\rm GeV})^2$.
However, as long as $D_S > -(260\>{\rm GeV})^2$, we find
$\sin^2 (\beta-\alpha) > 0.9 $ in this example.

\noindent
Case C: In Figure 3, we again plot $\mel$, $\mer$, $m_h$ and $m_A$,
now as a function of $\dnu=D_S$ with $D_Y=0$ and all other parameters
fixed as in the template model. In this case, nothing very remarkable
happens to the Higgs masses or couplings over the entire allowed
range for $\dnu=D_S$. Both $m_A$ and $m_h$ fall somewhat as $\dnu=D_S$
increases, but $\sin^2(\beta-\alpha) > 0.96$ always in this example.
The lower bound on $\dnu=D_S$ is set by the sneutrino contribution to
the invisible width of the $Z$ boson
(or, equivalently, by $\mel > 82$ GeV
for this value of $\tan\beta$). The upper bound on $\dnu=D_S$ is set by
the lower limit of 45 GeV on $\mer$. Note that for
$\dnu=D_S $ less than only $ -(59\> {\rm GeV})^2$, one finds $\mer > \mel$, in
contradistinction to the usual situation in models with universal
boundary conditions.

The rate for Higgs scalar boson production from $e^+e^- \rightarrow Zh$
is equal to $\sin^2(\beta-\alpha)$ times the corresponding rate for a
Standard Model Higgs scalar boson of the same mass.
The situation (illustrated by cases A and B) of light pseudoscalar masses
together with small values of $\sin^2(\beta-\alpha)$ can occur for all
values of $\tan\beta$ and $m_{\rm top}$ when appropriate D-terms are
present, unlike in the case of universal boundary conditions. For all
values of $\tan\beta$, we found some models for which $\sin^2(\beta-\alpha)$
was very close to zero. This generally
happens near the ``edge" of D-term parameter space, i.e., for values of
$D_Y$, $\dnu$, and $D_S$ near their limits. When this occurs, the
production cross-section for $h$ is reduced, but the rate for $hA$ production
is correspondingly enhanced, so that $hA$ production may be observable
at LEP2.

We now consider the possibility of arbitrary values for all of the
parameters. The remaining results of this section were obtained by
an exhaustive exploration of the parameter space, with the following
constraints:
$$
\eqalign{
-(200\>{\rm GeV})^2 < &\>m_0^2 < (800\>{\rm GeV})^2;\cr
40\>{\rm GeV} < &\>m_{1/2} < 400\>{\rm GeV};\cr
1.5 < \tan\beta < 60;\qquad & \qquad {\rm sign}(\mu) = \pm ;\cr
160 \>{\rm GeV} < &\> m_{\rm top} < 190\>{\rm GeV};\cr
-(500\>{\rm GeV})^2 < D_Y,&\dnu,D_S < (500\>{\rm GeV})^2\> .\cr
}\eqno(runlimits)
$$
As explained in the previous section, we allow for the possibility of
negative $m_0^2$. (We have checked, however, that this possibility does
not have a strong impact on the general results outlined below,
although it clearly can affect physics within individual models.) We
impose constraints on negative squark and slepton running (mass)$^2$ terms at
$ M_I = 10^{10}$ GeV corresponding to each of the flat directions
$({\tilde u}, {\tilde e})$;
$({\tilde u}, {\tilde d})$;
$({\tilde Q}, {\tilde L})$;
$({\tilde d}, {\tilde L})$; and
$({\tilde e}, {\tilde L})$ (in each case with appropriate flavor structure).
Here $({\tilde u}, {\tilde e})$
denotes for example the D-flat and F-flat direction
$$
\tilde u_R = \pmatrix{v \cr 0 \cr 0}
\qquad\tilde c_R = \pmatrix{0 \cr v \cr 0}
\qquad\tilde t_R = \pmatrix{0 \cr 0 \cr v}
\qquad\tilde \tau_R = \sqrt{2} v\>,
$$
which implies the constraint
$ 2 m^2_{\tilde u_R} + m^2_{\tilde t_R}
+ 2 m_{\tilde \tau_R}^2 > 0$ at $M_I$.

The range allowed for $A_0$ was determined for each model by the requirement
that
there be no color-breaking global minima of the scalar potential
at the electroweak scale in the D-flat (but not F-flat) directions
$\langle {\tilde t}_L \rangle = \langle {\tilde t}_R \rangle =
\langle H_u \rangle \not=0$ and
$\langle {\tilde b}_L \rangle = \langle {\tilde b}_R \rangle =
\langle H_d \rangle \not=0$. In addition, we required that there not
be global color-breaking minima in the non-D-flat directions
$\langle {\tilde t}_L \rangle \not=0$ or
$\langle {\tilde t}_R \rangle \not=0$, in each case with all other
VEVs vanishing. The latter requirements amount to approximately
$$
\eqalignno{
&m^2_{\tilde t_L} > - M_Z^2 |\cos (2\beta )|
\left ( {\alpha_3 + 3 \alpha_2/4
\over 3 \alpha_2 + 9 \alpha_1/5} \right )^{1/2}
 &(stopconl) \cr
&m^2_{\tilde t_R} > - M_Z^2 |\cos (2\beta )|
\left ( {\alpha_3 + 4 \alpha_1/5
\over 3 \alpha_2 + 9 \alpha_1/5} \right )^{1/2}
 &(stopconr) \cr
}
$$
respectively at the weak scale.
Here $m^2_{\tilde t_L}$ and $m^2_{\tilde t_R}$ are
the running stop mass parameters in the absence of electroweak breaking,
which appear in the usual stop mass matrix as
$$
\pmatrix{m^2_{\tilde t_L} + m_{\rm top}^2 - .35 |\cos 2\beta | M_Z^2
& m_{\rm top} (A_t + \mu \cot\beta) \cr m_{\rm top} (A_t + \mu \cot\beta)
& m^2_{\tilde t_R} + m_{\rm top}^2 - .15 |\cos 2\beta | M_Z^2 \cr
}\> .
$$
The constraints \(stopconl) and \(stopconr) effectively eliminate the
possibility of a stop squark much lighter than $m_{\rm top}$
with a small stop mixing angle, and thus
limit certain combinations of the D-terms. In this study we do
not allow the $K_i$ introduced in the previous section to vary because of
the practical need to confine the scope of the investigations, although
it would certainly be interesting to consider this.
We should remark that the limits
on the dimensionful parameters in \(runlimits) were motivated both by
practical considerations (e.g.~computer running time) and our theoretical
prejudices. Some of the results below would have to be generalized if,
for example, we allowed $m_0^2 < -(200\>{\rm GeV})^2$, or larger
D-terms. With that caveat, we now proceed to see how the allowed
supersymmetric spectrum is modified by the presence of D-terms.

Let us first consider the fate of some correlations between sparticle masses
which may be taken for granted in the case of universal boundary conditions.

In Figure 4, we show the allowed region in the
$(m_{\tilde e_L}, m_{\tilde e_R})$ plane for all
models with universal boundary conditions (bounded by solid lines),
and the additional region (bounded by dashed lines)
allowed for all models with
arbitrary D-terms. In the universal case, viable models lie within a
fairly restricted wedge; in particular, $m_{\tilde e_L} \geq m_{\tilde e_R}$
holds as a general rule. (If we relaxed the restriction $m_{1/2} < 400$ GeV
in our original choice of parameter space, the  region allowed by
universal boundary conditions would be enlarged for $\mel > 280$ GeV.)
With D-terms present, $\mer$ and $\mel$ can each
independently essentially saturate their lower limits.
The lower limit on $\mel$ in all models considered here
is about 75 GeV; except for small $\tan\beta$, this limit comes from
the sneutrino contribution to the invisible width of the $Z$
and the sum rule \(elersumrule).
As we saw from our examples A, B, and C, the slepton masses often
(but certainly not always) set the limits on the D-terms.

Figure 5 depicts in the same way the allowed regions in the
$(\mel,\mul)$ plane. Here the correlation in the universal boundary
condition case is looser,
because $\mul$ grows much faster with $m_{1/2}$ than $\mel$ does.  In the
presence of D-terms the
allowed region is considerably larger, but not quite enough
to saturate the experimental lower limits on squark masses when
$\mel > 150$ GeV. Still, a
substantial area is opened up for which $\mul < \mel$, a situation
which does not occur in the case of universal boundary conditions. A
similar inversion of the usual mass hierarchy can occur for each of the
other pairs of squarks and sleptons, depending on the values of
$D_Y$, $\dnu$, and $D_S$.

The masses of the first and second family squarks are always
very strongly correlated
with each other in the case of universal boundary conditions. In Figure
6, we show the (extremely narrow) allowed region in the
$(\mur,\mdr)$ plane with solid lines, and the much larger region allowed
by D-term contributions is enclosed in the dashed lines.
However, the correlation between the two squark masses is not completely
destroyed by D-terms,
in the sense that e.g.~$\mur < 200$ GeV implies $\mdr < 300$
GeV; the converse is not true, however. The squark masses do not always
saturate their lower bounds from direct searches (although $\mdr$ can be
quite small), mainly because the relative magnitude of the D-terms is
constrained by the lower limits on the slepton masses, by Higgs
boson searches, and by correct electroweak symmetry breaking.

Figure 7 depicts the allowed regions in the $(\mur,\mdl)$ plane. The
wedge (bounded by solid lines)
allowed by universal boundary conditions is slightly wider than in Figure 6.
This time the correlation between the squark masses is much stronger
than in Figure 6,  even for D-terms as large as allowed by other constraints.
This can be easily understood from \(mssmds); since $\mur^2$ and $\mdl^2$
obtain the same contribution from $\dnu$ and $D_S$,
a big difference between them can only occur if $D_Y$ dominates. However,
such a situation is limited by the lower bound on $\mer$ (if $D_Y >0$)
or by the sneutrino contribution to the invisible width of the $Z$ boson (if
$D_Y< 0$). There is an interesting region for which
$\mur > \mdl,\mul$; in this case the number of charginos  and
second-lightest neutralinos  from
gluino decays might be substantially increased over the usual
expectation from universal boundary conditions. This can also occur if
$\mdr > \mdl,\mul$.

The mass of $\tilde d_R$ (and $\tilde b_1$) can even be much less than
the lightest stop when D-terms are present.
In Figure 8, we show the allowed regions in the
$(\mdr,m_{\tilde t_1})$ plane. It should also be noted that with
D-terms present, $\tilde d_R$ especially can be considerably lighter than
the gluino, a sharp contrast from the usual rule in the case of universal
boundary conditions that all first and second
family squark masses are greater than about .85 $M_{\tilde g}$.

As remarked in section 2, large D-terms can also have a significant
effect on the neutralino and chargino sectors, through their impact on the
$\mu$ parameter as implied by \(changemu). In much of the allowed parameter
space in the MSSM with universal boundary
conditions, $\mu$ is large compared to $M_2$ so that the LSP ($\tilde N_1$),
the second-lightest neutralino ($\tilde N_2$),
and the lightest chargino ($\tilde C_1$) are usually gaugino-like.
In Figure 9, we show the allowed regions in the $(\mu,M_2)$ plane for
universal scalar masses (within solid lines) and for non-zero D-terms (within
dashed lines). In the latter case there is a substantial new region
with relatively smaller $\mu$. The region with $\mu \gg M_2$ is also enlarged.
If we restricted ourselves to any particular fixed values of $\tan \beta$
and $m_{\rm top}$, the enlargement of the allowed region in the
$(\mu,M_2)$ plane would be even more dramatic.
We find that eq.~\(changemu) is usually a quite good indicator of the
effect of the D-terms on $\mu$, even with full one-loop radiative corrections
in the Higgs sector taken into account.

In the MSSM, there is a well-known correlation between the masses of the
lightest chargino and the second-lightest neutralino.
In the limit that $M_Z /(\mu \pm M_{1,2})$ is small, one finds[\cite{mr}]
$$
m_{\tilde C_1} \approx m_{\tilde N_2 } \approx
 M_2 - {M_W^2 (M_2 + \mu \sin 2 \beta) \over
\mu^2 - M_2^2}
$$
However, when $|\mu | \approx M_2$, this expansion fails so that one
might expect a rather different relation between these masses. In Figure 10
we show the allowed regions in the $(m_{\tilde C_1}, m_{\tilde N_2})$ plane
for universal and for D-term boundary conditions.
The latter region is significantly larger, particularly for smaller
$\mu,M_2$. On the other hand, we find that other important features of the
chargino and neutralino sector are largely unchanged by the presence of
D-terms. An important example concerns the mass differences
$m_{\tilde C_1} - m_{\tilde N_1}$ and
$m_{\tilde N_2} - m_{\tilde N_1}$
which are important in determining the visibility
of a possible trilepton signal for SUSY
at a hadron collider, or of chargino pair-production at an $e^+e^-$ collider.
We find that $m_{\tilde C_1} - m_{\tilde N_1} > 15$ GeV
always within our parameter space \(runlimits), even when $|\mu |$ is
lowered with respect to $M_2$ by D-terms. The allowed regions in the
$(m_{\tilde N_1}, m_{\tilde C_1} )$ plane are shown in Fig.~11 for
universal (enclosed in solid lines)
and D-term (enclosed in dashed lines) boundary conditions.
The charged lepton coming from chargino decay
$\tilde C_1 \rightarrow \tilde N_1 l^+ \nu$ should thus
always be energetic enough to serve as a trigger,
if the corresponding branching ratio is large enough.
The mass difference
$m_{\tilde N_2} - m_{\tilde N_1}$ is also always larger than 7 GeV
(and in the vast majority of models we found it was larger than 15 GeV).
In Figure 12 we show the allowed regions in the
$(m_{\tilde N_1}, m_{\tilde N_2} )$ plane.
Of course, these results  are really
more of a statement about the form of the neutralino and chargino mass
matrices (assuming gaugino mass unification)
after all present experimental constraints have been imposed than it
is about the consequences of D-terms, which only affect this sector indirectly
through their effect on $\mu$.

[In the case that $\mu \ll m_{1/2}$, one could imagine having almost
degenerate $\tilde C_1$, $\tilde N_2$, and $\tilde N_1$ with large higgsino
content. While this is a logical possibility, it is
impossible to achieve in practice without extreme fine-tuning of parameters
including $D_Y$, $\dnu$, and $D_S$. In particular, it did not occur in
the search of parameter space described above. This situation would also
imply that the relic density of LSP's would be
far too small for them to constitute the cold dark matter.]
{\phantom{\cite{rblep}\cite{bf}\cite{wkk}\cite{cw}\cite{gjs}}}

We conclude this section by mentioning the effect of D-terms on the
$Z$-peak observable $R_b = \Gamma(Z
\rightarrow b\overline b) / \Gamma (Z\rightarrow {\rm hadrons})$. The
Standard Model prediction for this quantity
is lower by about 2$\sigma$ than the
experimental value $R_b = 0.2202 \pm .0020$ obtained at LEP[\cite{rblep}].
Supersymmetric corrections to $R_b$ are almost always tiny
[\cite{bf}-\cite{gjs}].
In particular, in the MSSM with universal
boundary conditions, $R_b$ never gets closer than 1.5$\sigma$ to
the central experimental value. In the most general supersymmetric
scenario, the prediction for $R_b$ can only be reconciled with
the 1$\sigma$ experimental range if certain strong conditions are met.
For $\tan\beta < 30$, these include light stops which are predominantly
$\tilde t_R$, and light charginos with a significant higgsino content.
One might imagine that these conditions could be attained in the presence
of D-terms. Somewhat surprisingly, we find that this is not the case;
an examination of viable models satisfying the constraints
\(runlimits) finds that the prediction for $R_b$
can only slightly exceed the values indicated in Figure 1 of
ref.~[\cite{wkk}], and never enter the 1$\sigma$
experimentally allowed range. This may be understood qualitatively
from the previous results; when the superpartners are near their experimental
lower bounds (as required for larger $R_b$), one finds that only relatively
mild deviations from universal boundary conditions
can occur. The extent to which the mass of the lightest stop squark
can be lowered without having a large stop mixing angle is limited by the
constraint \(stopconr) on color-breaking minima. On the other hand,
a large stop mixing angle generally implies that $|\mu|$ is relatively
large (so that the lighter chargino is not higgsino-like) or that $A_t$ is
large (which is problematic because of possible color breaking minima of the
scalar potential in a D-flat direction).
This implies that the experimental value of $R_b$ must
fall if minimal SUSY is correct with universal gaugino mass boundary
conditions and only D-term contributions to scalar mass non-universality
as imposed here.

\subhead{4. Conclusion}
\taghead{4.}

In this paper, we have examined the possibility of D-term contributions
to soft scalar masses in the MSSM which arise if the unbroken
gauge group at very high energies is an arbitrary subgroup of $E_6$. These
contributions are parameterized by three quantities $D_Y$, $D_X$, and $D_S$.
The effect of one-loop
RG evolution on these parameters can always be absorbed into
a redefinition of $D_Y$. The sizes of $D_Y$, $D_X$, and $D_S$ (for a given
set of other parameters) are limited by direct and indirect constraints on
sparticles (especially sleptons) and Higgs scalar boson masses, as well
as from correct symmetry breaking.

Of course, there is no guarantee that the D-terms will be large enough to
make their presence felt in comparison to the universal contributions to
scalar masses from $m_0^2$ and gaugino loops. Furthermore, it is clear
that a future observation of apparent scalar mass non-universality at $M_U$
may not be unambiguously ascribed to D-terms, without further theoretical
input. Still, checks of relationships which hold in the MSSM between squark
and slepton masses will provide an important clue to physics at very high
energies. With experimental constraints imposed, we found that the impact of
the D-terms might be most dramatic on squark and slepton mass relations [see,
for example, eq.~\(sumruleb) and Fig.~7].
In several important respects, we found that deviations from the universal
boundary conditions tend to be rather mild, especially when superpartners
are relatively light. For example, we did not succeed
in finding a viable model with non-zero D-terms for which $R_b$ was within
1$\sigma$ of the present experimental value. We did find models at all
values of $\tan\beta$ and $m_{\rm top}$
for which $\sin^2(\beta-\alpha)$ was close to zero,
in which case Higgs production through the usual Standard Model mechanism
$e^+e^- \rightarrow Zh$ will be impossible to detect.
However, this generally occurred only when the D-terms were
quite close to their experimental bounds, except for large $\tan\beta$.
If scalar mass non-universality
proves to be a phenomenological necessity, D-term contributions will be
an attractive theoretical candidate, since they maintain the natural
suppression of flavor-changing neutral currents.

\noindent Acknowledgments:
We are grateful to Manuel Drees, Gordon Kane, Riccardo Rattazzi,
and James Wells for their help.
This work was supported in part by the U.S. Department of Energy.

\references

\refis{supergravity} A.~Chamseddine, R.~Arnowitt, and P.~Nath,
\prl 49, 970, 1982; H.~P.~Nilles, \pl 115B, 193, 1982;
L.~E.~Ib\'a\~nez, \pl 118B, 73, 1982;
R.~Barbieri, S.~Ferrara and C.~Savoy, \pl 119B, 343, 1982;
L.~Hall, J.~Lykken and S.~Weinberg, \pr D27, 2359, 1983;
P.~Nath, R.~Arnowitt and A.~Chamseddine, \np B227, 121, 1983.

\refis{kmy} Y.~Kawamura, H.~Murayama, and M.~Yamaguchi, \pl B324, 52, 1994;
\pr D51, 1337, 1995.
%``Low-Energy Effective Lagrangian in Unified
%Theories with Non-Universal Supersymmetry Breaking Terms", preprint
%LBL-35731, hep-ph/9406245.

\refis{rr} G.~G.~Ross and R.~G.~Roberts, \np B377, 571, 1992.

\refis{an} R.~Arnowitt and P.~Nath, \prl 69, 725, 1992.

\refis{dn} M.~Drees and M.~M.~Nojiri, \pr D45, 2482, 1992.

\refis{mr} S.~P.~Martin and P.~Ramond, \pr D48, 5365, 1993.

\refis{hk} J.~S.~Hagelin and S.~Kelley, \np B342, 95, 1990.

\refis{ch} H.-C. Cheng and L.~J.~Hall, ``Squark and slepton mass relations
in grand unified theories", preprint LBL-35950, hep-ph/9411276.

\refis{wkk} J.~D.~Wells, C.~Kolda, and G.~L.~Kane, \pl B338, 219, 1994.

\refis{mn} D.~Matalliotakis and H.~P.~Nilles, \np B435, 115, 1995.
%``Implications of
%non-universality of soft terms in supersymmetric grand unified theories",
%preprint TUM-HEP-201/94, hep-ph/9407251.

\refis{kkrw} G.~L.~Kane, C.~Kolda, L.~Roszkowski, and J.~Wells,
\pr D49, 6173, 1994.

\refis{cpr} D.~J.~Casta\~no, E.~J.~Piard, and P.~Ramond,
\pr D47, 232, 1992.

\refis{op} M.~Olechowski and S.~Pokorski, \pl B344, 201, 1995.
%``Electroweak symmetry breaking with nonuniversal scalar
%soft terms and large $\tan \beta$ solutions", preprint MPl-PHT-94-40
%hep-ph/9407404.

\refis{op1} M.~Olechowski and S.~Pokorski, \np B404, 590, 1994.

\refis{rsh} R.~Rattazzi, U.~Sarid, and L.~J.~Hall,
``Yukawa unification: the good, the bad and the ugly",
preprint SU-ITP-94-15, hep-ph/9405313.

\refis{witten} E.~Witten, \np B258, 75, 1985.

\refis{fhkn} A.~E.~Faraggi, J.~S.~Hagelin, S.~Kelley, and
D.~V.~Nanopoulos, \pr D45, 3272, 1992.

\refis{pp} N.~Polonsky and A.~Pomarol, \prl 73, 2292, 1994;
``Non-universal GUT corrections to the soft terms and their
implications in supergravity models", preprint UPR-0627T,
hep-ph/9410231.

\refis{drees} M.~Drees, \pl B181, 279, 1986.

\refis{llm} A.~Lleyda and C.~Mu\~noz, \pl B317, 82, 1993.

\refis{unification} P.~Langacker, in Proceedings of the
PASCOS90 Symposium, Eds.~P.~Nath
and S.~Reucroft, (World Scientific, Singapore 1990)
J.~Ellis, S.~Kelley, and D.~Nanopoulos, \pl B260, 131, 1991;
U.~Amaldi, W.~de Boer, and H.~Furstenau, \pl B260, 447, 1991;
P.~Langacker and M.~Luo, \pr D44, 817, 1991.

\refis{reviews}
For reviews, see H.~P.~Nilles,  \prpts 110, 1, 1984 and
H.~E.~Haber and G.~L.~Kane, \prpts 117, 75, 1985.

\refis{il} L.~E.~Ib\'a\~nez and D.~Lust, \np B382, 305, 1991.

\refis{sw} S.~K.~Soni and H.~A.~Weldon, \pl 126B, 215, 1983.

\refis{bbo} V.~Barger and M.~S.~Berger and P.~Ohmann, \pr D49,
4908, 1994.

\refis{copw} M.~Carena, M.~Olechowski, S.~Pokorski, and C.~Wagner,
\np B419, 213, 1994.

\refis{br} P.~Binetruy and P.~Ramond, ``Yukawa Textures and Anomalies",
preprint hep-ph/9412385.

\refis{bf} M.~Boulware and D.~Finnell, \pr D44, 2054, 1991.

\refis{chsw} P.~Candelas, G.~T.~Horowitz, A.~Strominger, and E.~Witten,
\np B258, 46, 1985.

\refis{dp} S.~Dimopoulos and A.~Pomarol, ``Non-Unified Sparticle and Particle
Masses in Unified Theories", preprint CERN-TH/95-44, hep-ph/9502397.

\refis{cw} M.~Carena and C.~Wagner, ``Higgs and Supersymmetric Particle
Signals at the Infrared Fixed Point of the Top Quark Mass", preprint
CERN-TH.7397/94, hep-ph/9408253

\refis{fpmt} J.~L.~Feng, M.~E.~Peskin, H.~Murayama, and X.~Tata,
``Testing Supersymmetry at the Next Linear Collider", preprint
SLAC-PUB-6654, hep-ph/9502260.

\refis{ff} J.~L.~Feng and D.~Finnell, \pr D49, 2369, 1994.

\refis{jlc1} JLC Group, ``JLC-I", KEK Report 92-16, Tsukuba, 1992.

\refis{pdb} Particle Data Group, L.~Montanet et.~al., \pr D50, 1173, 1994.

\refis{rblep} The LEP collaborations, preprint CERN/PPE/94-187.

\refis{gjs} D.~Garcia, R.~A.~Jim\'enez, and J.~Sol\`a, ``Full electroweak
quantum effects on $R_b$ in the MSSM", preprint UAB-FT-344, hep-ph/9410311.

\refis{kt} Y.~Kawamura and M.~Tanaka,
\journal Prog. Theor. Phys., 91, 949, 1994.

\endreferences
%\vfill\eject
\doublespace
\subhead{Figure Captions}

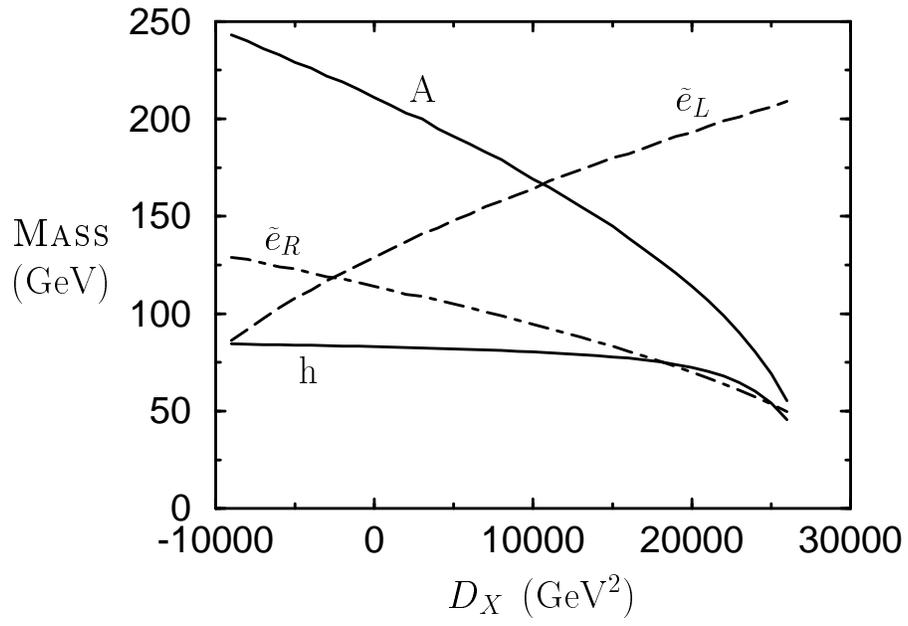
\noindent{} Figure 1: The masses of the lightest Higgs scalar boson $h$ (lower
solid line), pseudoscalar Higgs boson $A$ (upper solid line),
left-handed selectron $\tilde e_L$ (dashed line), and right-handed selectron
$\tilde e_R$ (dot-dashed line), as a function of the parameter $D_X$
added to the template model described in the text.

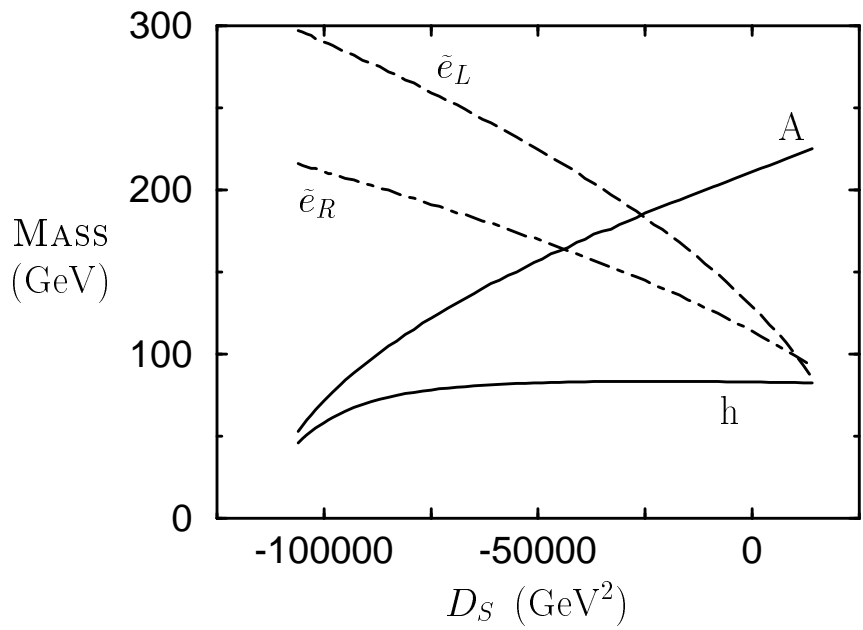
\noindent{} Figure 2: The same quantities are plotted as in Figure 1, but now
 as a function of $D_S$ added to the template model described in the text.

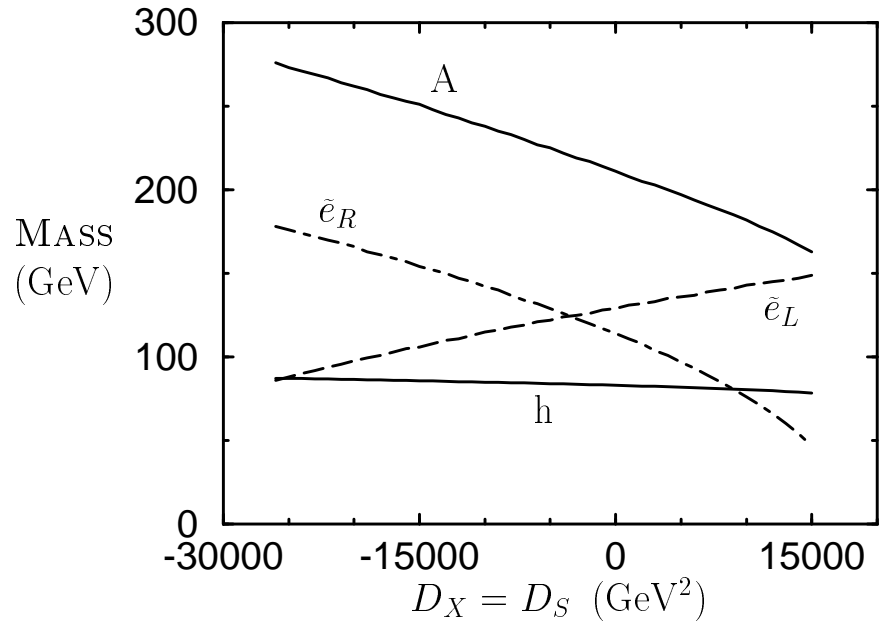
\noindent{} Figure 3: The same quantities are plotted as in Figure 1, but now
as a function of $D_X=D_S$ added to the template model described in the text.

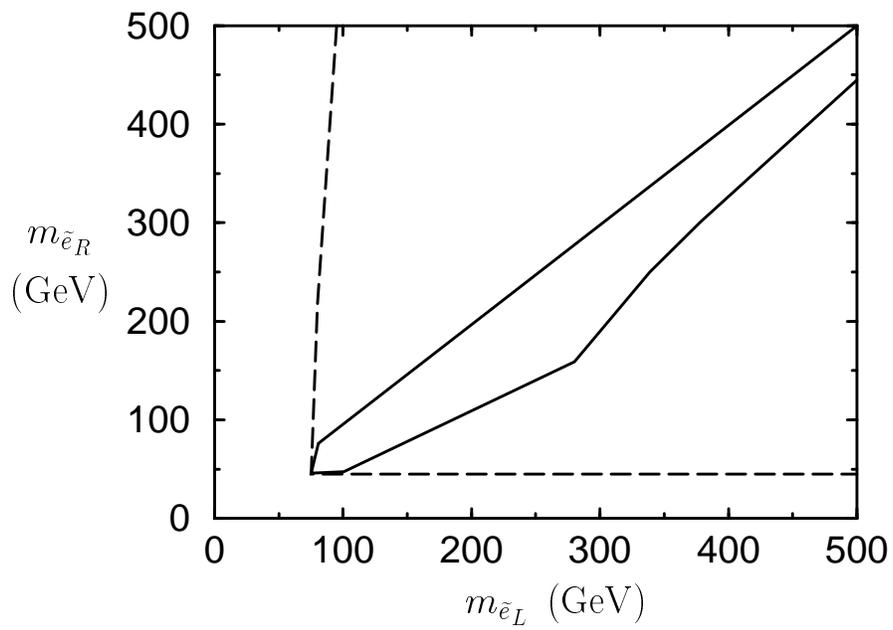
\noindent{} Figure 4: The allowed regions in the $(\mel,\mer)$ plane
for universal boundary conditions (inside solid lines) and arbitrary
D-term boundary conditions (inside dashed lines),
for models satisfying \(runlimits).

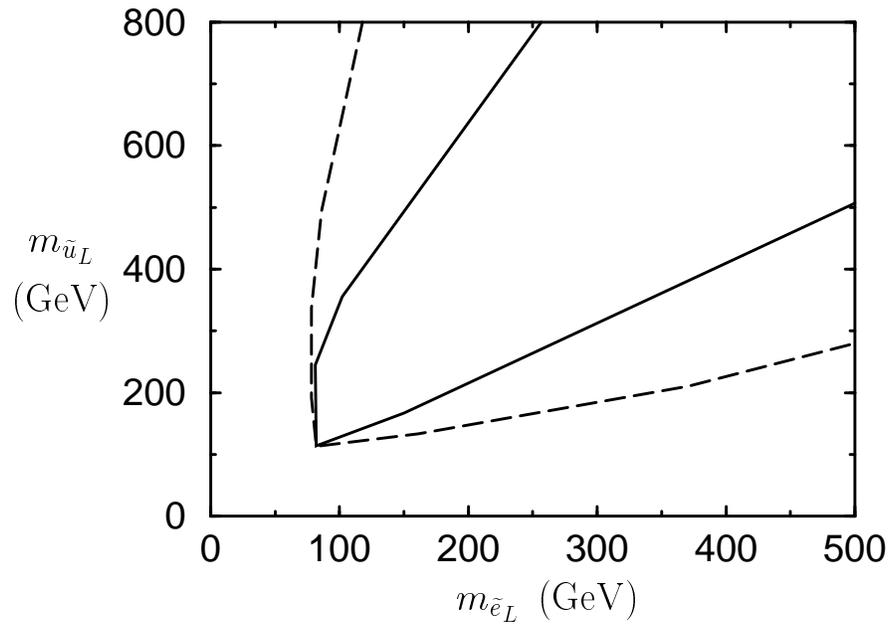
\noindent{} Figure 5: The allowed regions in the $(\mel,\mul)$ plane.

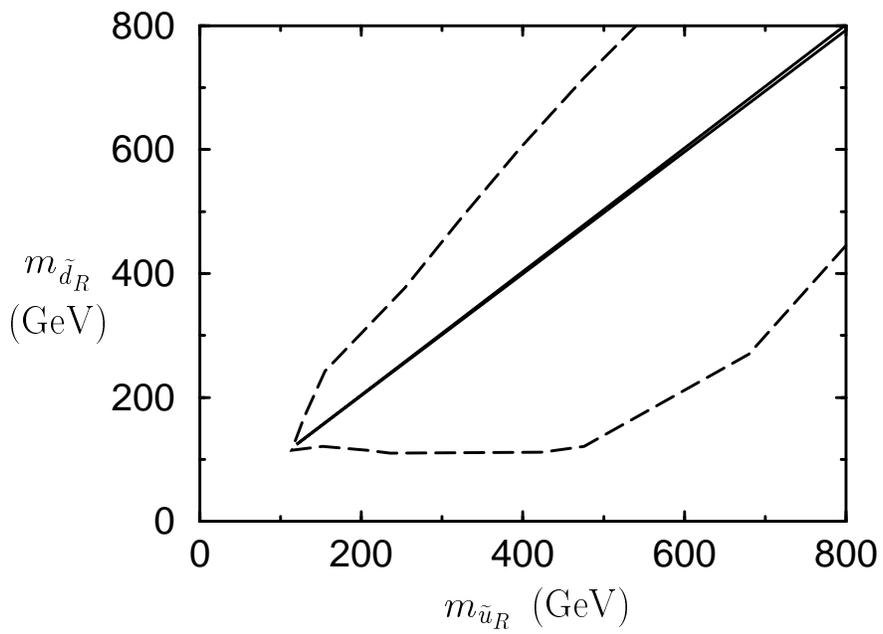
\noindent{} Figure 6: The allowed regions in the $(\mur,\mdr)$ plane.

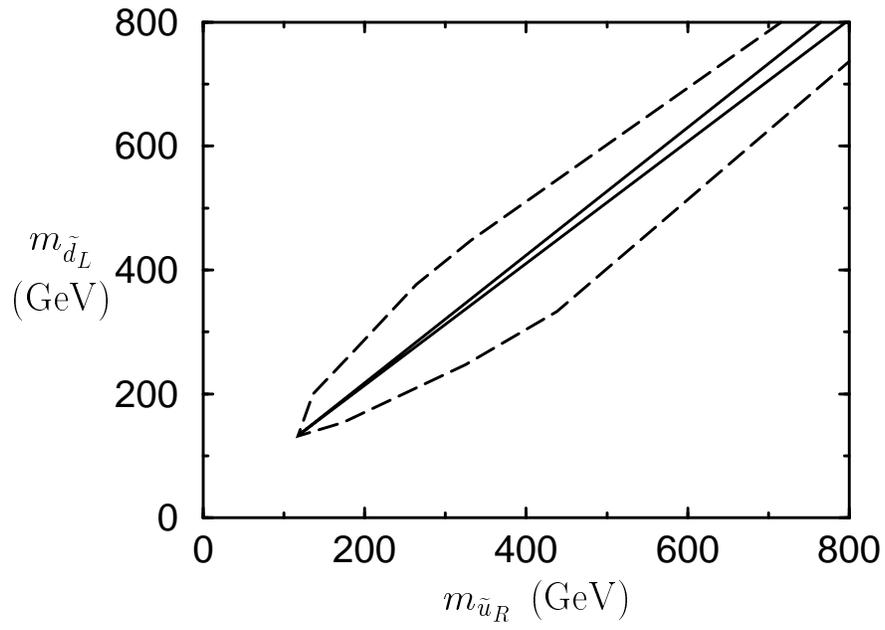
\noindent{} Figure 7: The allowed regions in the $(\mur,\mdl)$ plane.

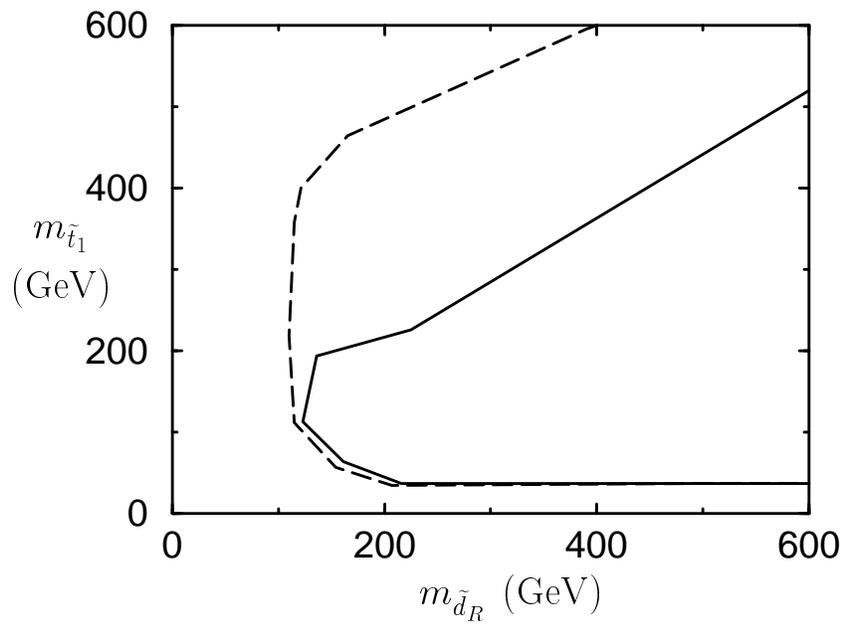
\noindent{} Figure 8: The allowed regions in the $(\mdr,m_{\tilde t_1})$ plane.

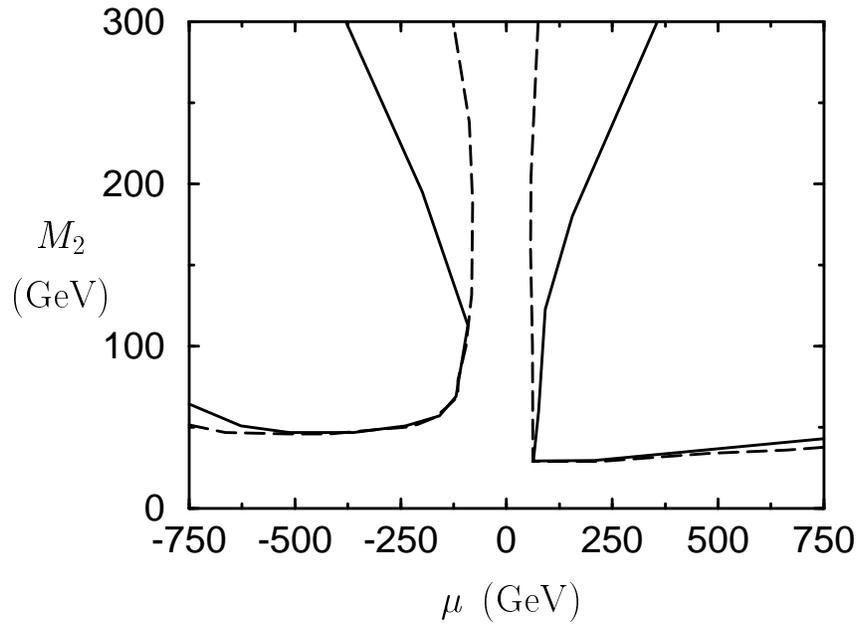
\noindent{} Figure 9: The allowed regions in the $(\mu,M_2)$ plane.

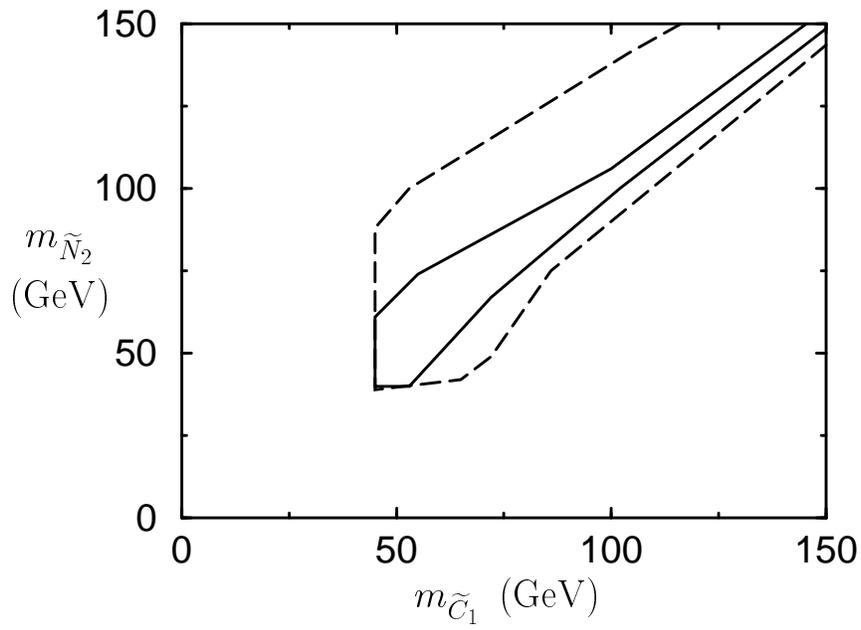
\noindent{} Figure 10: The allowed regions in the $(m_{\tilde C_1},
m_{\tilde N_2})$ plane.

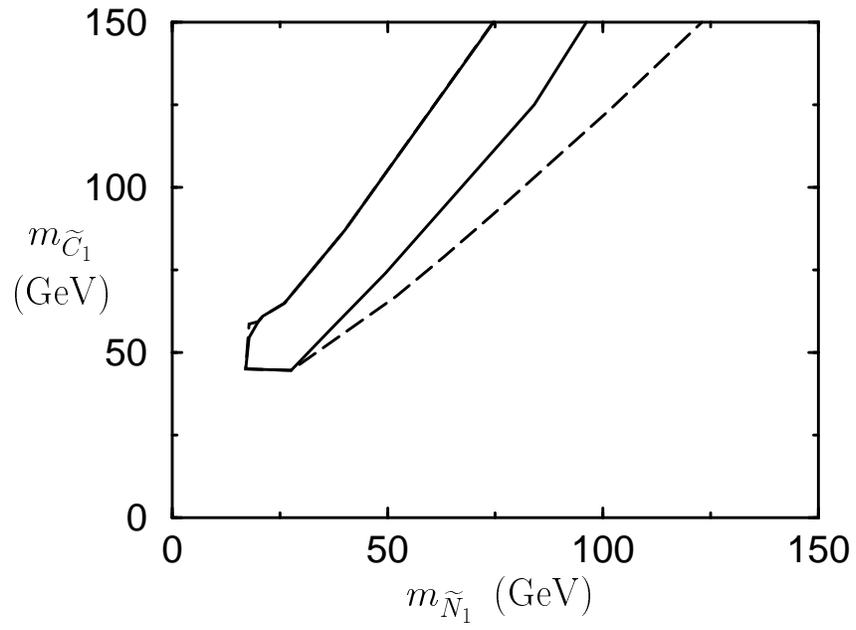
\noindent{} Figure 11: The allowed regions in the $(m_{\tilde N_1},
m_{\tilde C_1})$ plane.

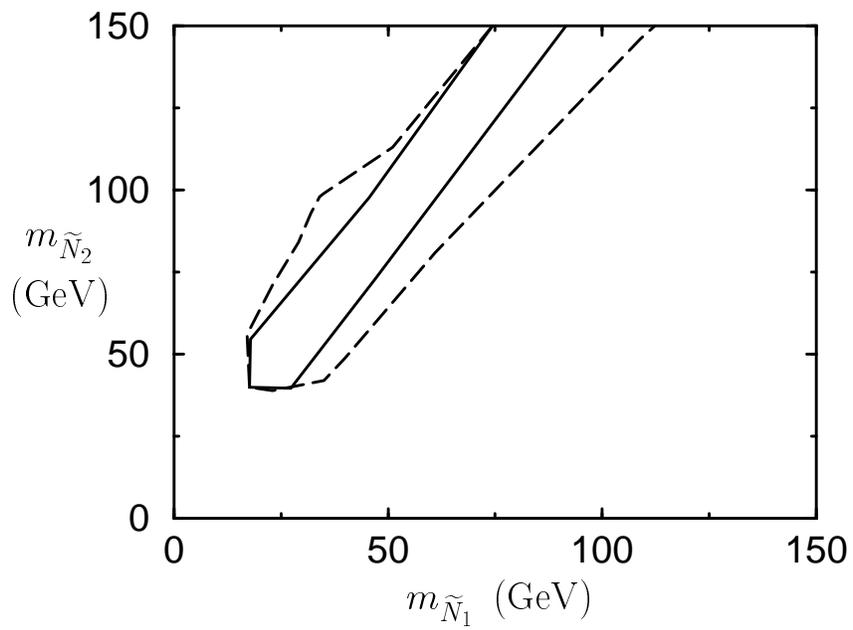
\noindent{} Figure 12: The allowed regions in the $(m_{\tilde N_1},
m_{\tilde N_2})$ plane.

\endit\end